\documentclass[journal]{IEEEtran}
\usepackage{amsfonts}
\IEEEoverridecommandlockouts

\ifCLASSINFOpdf
\else
\fi
\usepackage{multirow}
\usepackage{cite}
\usepackage{color}
\usepackage{graphicx}
\usepackage{amssymb }
\usepackage{algorithm}
\usepackage{algorithmic}
\usepackage{booktabs}
\usepackage{amsmath,bm}
\usepackage{colortbl}
\usepackage{makecell}
\usepackage[colorlinks]{hyperref}
\usepackage{amsthm}
\usepackage{tabularx}
\usepackage{array}
\usepackage{makecell}

\newtheorem{lemma}{Lemma}

\newtheorem{definition}{Definition}

\usepackage[percent]{overpic}
\begin{document}
\title{Energy-Accumulate-Then-Act Backscatter Communications With Retransmission Under Imperfect Feedback Detection}

\author{Rui Xu,
            Yinghui Ye,~\IEEEmembership{Member,~IEEE},
            Guangyue Lu,
            Liqin Shi,
            and
            Gan Zheng,~\IEEEmembership{Fellow,~IEEE}
            \vspace{-8mm}

\thanks{R. Xu, Y. Ye, G. Lu, and L. Shi are with the Shaanxi Key Laboratory of
Information Communication Network and Security, Xi'an University of Posts
\& Telecommunications, China (e-mails: $\rm dhlscxr@126.com$; $\rm connectyyh@126.com$; $\rm tonylugy@163.com$; $\rm liqinshi@hotmail.com$). }
\thanks{G. Zheng is with the School of Engineering, University of Warwick, CV4
7AL Coventry, U.K. (e-mail: $\rm gan.zheng@warwick.ac.uk$).}
}

\markboth{}
{Shi\MakeLowercase{\textit{et al.}}:}
\maketitle
\begin{abstract}
In this paper, we propose a threshold-triggered energy-accumulate-then-act backscatter communication system with retransmission under imperfect feedback detection.
Under this system, harvested energy is accumulated across transmission blocks, and the backscatter device switches among different operation modes, including energy harvesting, initial transmission, and retransmission, according to its stored energy and decoded feedback. To characterize the resulting dynamics,
the joint evolution of the packet-buffer and stored-energy states is modeled, based on which an embedded Markov chain is constructed to derive the steady-state energy distribution at the beginning of each update cycle.
Subsequently, the average age of information (AAoI) is derived from the statistics of the inter-update interval and its correlation with the preceding packet system time, while the packet loss rate (PLR) caused by imperfect feedback detection is also obtained.
Simulation results validate the analytical results and reveal that retransmission reduces PLR but may either improve or degrade AAoI, depending on packet-generation energy, data transmission rate, and feedback power.
 Moreover, lower feedback power shrinks the AAoI-favorable region of retransmission at low transmission rate, whereas the opposite trend emerges at higher rate.
\end{abstract}
\vspace{-0.5em}

\IEEEpeerreviewmaketitle

\vspace{-0.5em}
\section{Introduction}
\IEEEPARstart{A}{mbient} Internet-of-Things (AIoT) has emerged as a promising solution for enabling ultra-low-power and energy-autonomous connectivity in massive-scale, low-rate IoT applications, with AIoT-related standardization initiated by the Third Generation Partnership Project (3GPP) in Release 19 \cite{11119230,11153767,3gpp38769}.
Backscatter communication (BackCom) is a key enabling technology for AIoT \cite{10980384,11020590,3gpp38848,3gpp38769}, where a backscatter device (BD) transmits the generated data by modulating and reflecting ambient radio-frequency (RF) signals while harvesting energy from these signals and storing the harvested energy in a finite-capacity capacitor to sustain its operations. Accordingly, the operating decisions of BD, such as data generation (DG), energy harvesting (EH),
and backscatter transmission (BT), are constrained by their time-varying energy states, which, in turn, evolve according to the operating decisions. Such tight coupling between energy
states and operating decisions makes energy management a critical challenge in BackCom\cite{11153767,11482890}.

Most studies on BackCom adopt a \emph{per-transmission-block energy management protocol}, where the BD harvests energy from a portion of the incident RF signals during the BT phase by adjusting the power reflection coefficient (PRC), possibly supplemented by a dedicated EH phase, and uses the harvested energy to support BT under the energy causality constraint, with any residual energy discarded at the end of each transmission block. Under this protocol, extensive studies have been conducted from the perspectives of resource allocation \cite{10197252,11365549,11514149,10702441,11175015} and performance analysis\cite{10643608,9866050,9813592,10778600}.
However, \emph{a fundamental limitation remains: the harvested energy cannot be accumulated across transmission blocks, which may result in inefficient energy utilization.}
In practical AIoT scenarios, the harvested energy within a single transmission block may be insufficient to support the required operations due to the limited RF power density and severe path loss, resulting in potential energy outages and energy waste. To alleviate it, the BD needs to accumulate harvested energy across multiple transmission blocks until the stored energy becomes sufficient to perform operations, whereas this has been largely neglected in existing studies \cite{10197252,11365549,11514149,10702441,11175015,10643608,9866050,9813592,10778600}. \emph{Incorporating energy accumulation (EA) may introduce new analytical challenges, as the energy evolution becomes tightly coupled with the operating decisions of the BD and can no longer be independently characterized under the per-block protocol, complicating energy management and performance analysis.}

To tackle these limitations, our prior work in \cite{11482890} proposed a reliability-guaranteed
energy management protocol that considers EA under finite-capacity energy storage
and models the energy evolution using a Markov chain.
Specifically, the energy causality constraint and quality-of-service (QoS) requirement of the BD were jointly considered to optimize the PRC. When the QoS cannot be guaranteed, the BD switches to a dedicated EH mode and accumulates the harvested energy for subsequent transmission, thereby improving transmission reliability and avoiding energy waste during outage stages.
Based on the Markov chain model, the coupled evolution between energy states and transmission decisions was characterized, and the probability that the BD operates in the dedicated EH mode was derived.
Although the optimal PRC designed in \cite{11482890} can improve transmission reliability, its reliance on accurate instantaneous channel state information (CSI) introduces considerable channel estimation overhead, which is difficult to achieve for low-complexity AIoT.
However, without accurate CSI, the BD device cannot predict whether the channel is good enough for successful BT, thus it has to accumulate energy to a certain threshold and then transmit. The transmission may fail when the channel is poor, wasting the carefully accumulated energy.
This raises an open question:
\emph{How can we achieve reliable BackCom without relying on perfect CSI while considering EA?}

To address this challenge, retransmission (RT) has been specified by 3GPP as a viable solution \cite{3gpp38769}, which has also been investigated in BackCom systems \cite{11481662,10623260} and demonstrated reliability improvements. Typically, the BD relies on acknowledgment (ACK) or negative acknowledgment (NACK) feedback to determine whether a packet should be retransmitted.
However, due to the hardware limitations of low-complexity AIoT devices, non-coherent detection with relatively high decoding error probability is typically employed, which may lead to imperfect feedback detection (FD) and erroneous RT decisions \cite{9440981,9626550}. Specifically, if an ACK is incorrectly detected as a NACK, the BD unnecessarily retransmits an already delivered packet, delaying subsequent fresh updates and potentially degrading information freshness.
Conversely, if a NACK is incorrectly decoded as an ACK, the BD prematurely discards an undelivered packet, resulting in packet loss.
These facts motivate us to investigate the average age of information (AAoI) and packet loss rate (PLR) in RT-assisted BackCom systems under imperfect FD.

To improve energy utilization and transmission reliability,
we consider a threshold-triggered EA-then-act protocol\footnote{Here, ``act'' is used rather than ``transmit'', since this paper considers multiple operations of the BD, including DG, EH, BT, and FD, all of which affect the stored-energy evolution.} for RT-assisted BackCom under imperfect FD, where the BD autonomously generates status-update packets\footnote{Unlike the existing studies where data packets are assumed to be always available for transmission, practical AIoT applications, such as environmental monitoring and status tracking, require devices to generate data packets through sensing operations \cite{7435327}.} and accumulates harvested energy across transmission blocks. However, the joint consideration of EA, RT, and imperfect FD makes the long-term performance analysis challenging. Specifically, the detected feedback determines whether the BD retains the current packet for RT or proceeds to generate a new update, while the stored energy determines whether the required operation can be performed or additional EH is needed. These decisions, in turn, affect the energy evolution and subsequent status-update process. To address these coupled dynamics, we develop an analytical framework based on an embedded Markov chain to characterize the long-term AAoI, while the PLR is derived by accounting for erroneous feedback decisions. The main contributions are summarized as follows.
\begin{itemize}
  \item We propose a threshold-triggered EA-then-act protocol that integrates cross-block EA, RT, and imperfect FD.
     Unlike conventional per-block energy management scheme \cite{10197252,11365549,11514149,10702441,11175015,10643608,9866050,9813592,10778600}, the proposed protocol allows the BD to accumulate energy across transmission blocks, improving energy utilization.
     Different from the QoS-guaranteed  energy management scheme with EA in\cite{11482890},
     the proposed protocol employs RT to improve transmission reliability without requiring accurate CSI at the BD.
      Moreover, imperfect FD is considered to account for erroneous RT decisions for low-complexity AIoT devices.
      Accordingly, the BD switches among EH, initial transmission (IT), and RT based on its stored energy, packet status, and detected feedback, and the joint evolution of the packet-buffer and stored-energy states is characterized.
  \item We develop an analytical framework to characterize the long-term information freshness and transmission reliability under the proposed protocol. An embedded Markov chain is constructed to characterize the energy-state evolution, based on which the steady-state energy distribution at successful update epochs is derived. Subsequently, the moments of the inter-update interval and the cross moment between the system time of a successfully delivered packet and the subsequent inter-update interval are derived to obtain an analytical expression for the AAoI. Moreover, the PLR is obtained by characterizing premature packet discarding caused by erroneous FD.
  \item Extensive simulations are conducted to validate the analytical results and evaluate the impacts of EA, RT, and imperfect FD on AAoI and PLR. The trade-off between information freshness and reliability is further investigated by identifying the AAoI-favorable regions of RT under different DG energies, DT rates, and feedback powers. Comparisons with existing benchmarks further demonstrate that the proposed scheme achieves competitive performance under more practical assumptions.
\end{itemize}

\vspace{-0.5em}
\section{System Model}\label{sec:System Model}
This paper considers a typical bistatic BackCom system consisting of an RF source (S), a BD, and a receiver (R).
All nodes are equipped with a single antenna, and the BD is equipped with an energy storage unit with finite capacity $B_{\rm{max}}$ to accumulate harvested energy across transmission blocks. The system operates in a block-based manner, where each transmission block has a duration of $T_b$.
We assume independent quasi-static Rayleigh fading for all wireless links, where the channel gains remain constant within each transmission block and vary independently across different blocks.
Specifically, the channel coefficients of the S$\to$BD, BD$\to$R, and S$\to$R links for DT, and R$\to$BD link for FD in the $m$-th transmission block are denoted by $f\left( m \right)$, $g\left( m \right)$, $h\left( m \right)$, and $g_1\left( m \right)$, respectively, which are subject to the large-scale path loss ${{L_i}}$ and small-scale fading ${\xi _i}\left( m \right)$, i.e., $i\left( m \right) = \sqrt {{L_i}} {\xi _i}\left( m \right)$, $i = \left\{ {f,g,h,g_1} \right\}$. Since the small-scale fading ${\xi _i}\left( m \right)$ follows independent and identically distributed circularly symmetric complex Gaussian distribution, i.e., ${\xi _i}\left( m \right) \sim \mathcal{CN}\left( {0,1} \right)$, the corresponding channel gains follow exponential distributions, i.e., ${\left| {i\left( m \right)} \right|^2} \sim \exp \left( {{\lambda _i}} \right)$ with ${\lambda _i} = {1 \mathord{\left/
 {\vphantom {1 {{L_i}}}} \right.
 \kern-\nulldelimiterspace} {{L_i}}}$.

%

The transmission process of the considered system can be described as follows. The BD transmits self-generated data packets to the R via BackCom.
After decoding the received signals, the R feeds back an ACK/NACK message to the BD. Then, the BD decodes the received feedback message, based on which it determines whether to retransmit the previously generated data packet or generate a new data packet for IT. It should be pointed out that imperfect FD at the BD may result in erroneous transmission decisions, and DG is only performed before IT. Moreover, the execution of each operation is constrained by the available stored energy at the beginning of each transmission block. If the stored energy is insufficient to support the required operation, the BD first harvests energy until the required energy level is reached.

In the above transmission process, there are five basic actions performed by the BD, including \emph{DG}, \emph{EH}, \emph{IT}, \emph{RT}, and \emph{FD}, which are described as follows.

\subsubsection{DG}
The BD generates a new data packet through a sensing operation, which incurs an energy consumption (EC) of ${E_{{\rm{DG}}}}$. The generated packet is buffered at the BD and, if RT is required, remains available for subsequent RTs until the BD removes it according to the decoded feedback. Hence, DG is performed only before each IT, whereas RT delivers the packet already retained in the buffer and therefore incurs no additional EC for DG. The duration of DG is assumed to be negligible compared with the duration of DT.

\subsubsection{EH}
When the available stored energy is insufficient to support the required operations in the upcoming transmission block, the BD harvests energy from the received RF signals until sufficient energy is accumulated. Specifically, the S transmits a carrier signal $s\left( m \right) \sim \mathcal {CN}\left( {0,1} \right)$ in the $m$-th transmission block with transmit power $P_{\rm{s}}$.
At the BD, the received signal is given by
${y_{{\rm{BD}}}}\left( m \right) = \sqrt {P_{\rm{s}}} f\left( m \right)s\left( m \right)$,
and thus the harvested energy over the $m$-th transmission block can be expressed as
${E_{\rm{H}}}\left( m \right) = \eta {P_{\rm{s}}}{\left| {f\left( m \right)} \right|^2}{T_b}$,
where $\eta  \in \left[ {0,1} \right]$ denotes the energy conversion efficiency.
Notably, the noise at the BD can be ignored following existing works \cite{10129192}.

\subsubsection{IT}
After generating a fresh packet, the BD transmits its signals via BackCom by riding on the received signal ${y_{{\rm{BD}}}}\left( m \right)$.
Applying a full-reflection mode for BackCom, i.e., the PRC is set to one, the received signal at the R is given by
\setlength{\abovedisplayskip}{2pt}
\setlength{\belowdisplayskip}{2pt}
      \begin{align}\label{1A}
{y_{\rm{R}}}\!\left(\! m\! \right) \!=\!\! \sqrt {P_{\rm{s}}} f\!\left( \!m \! \right)g\!\left( \!m \!\right)s\!\left( \!m \!\right)c\!\left( \!m\! \right)
 \!+ \!\!\sqrt {P_{\rm{s}}} h\!\left( \!m\! \right)s\!\left(\! m \!\right) \!+\! \!w\!\left( \!m \!\right),
\end{align}
      where $c\left( m \right) \sim \mathcal{CN}\left( {0,1} \right)$ denotes the transmitted signal of status update at the BD in the $m$-th transmission block, $w\left( m \right) \sim \mathcal{CN}\left( {0,{\sigma ^2}} \right)$ is the additive white Gaussian noise (AWGN) at the R.
       Since the carrier signal $s\left( m \right)$ is known for the R and the CSI of $h\left( m \right)$ is assumed to be available, the direct-link
       interference $\sqrt {P_{\rm{s}}} h\left( m \right)s\left( m \right)$ can be perfectly removed via successive interference cancellation before decoding $c\left( m \right)$ at the R. Accordingly, the decoding signal-to-noise ratio (SNR) for $c\left( m \right)$ at the R is
       $\gamma\! \left( m \right) \!\!= \!\! {{{P_{\rm{s}}}{{\left| {f\!\left( m \right)} \right|}^2}{{\left| {g\!\left( m \right)} \right|}^2}} \mathord{\left/
 {\vphantom {{P{{\left| {f\!\left( m \right)} \right|}^2}{{\left| {g\!\left( m \right)} \right|}^2}} \!\!{{\sigma ^2}}}} \right.
 \kern-\nulldelimiterspace} \!\!{{\sigma ^2}}}$.
 Thus, the outage probability (OP) of the $m$-th transmission block is given by
 \setlength{\abovedisplayskip}{2pt}
\setlength{\belowdisplayskip}{2pt}
 \begin{align}\label{2A}
 {P_{{\rm{out}}}} \!=\!\Pr \!\left\{ {\gamma \!\left( m \right) \!<\! {\gamma _{{\rm{th}}}}} \!\right\} \!=\! 1 \!-\! 2\sqrt {{\lambda _f}{\lambda _g}z} {K_1}\!\left( {2\sqrt {{\lambda _f}{\lambda _g}z} } \right),
\end{align}
where $z = {{{\sigma ^2}{\gamma _{{\rm{th}}}}} \mathord{\left/
 {\vphantom {{{\sigma ^2}{\gamma _{{\rm{th}}}}} {P_{\rm{s}}}}} \right.
 \kern-\nulldelimiterspace} {P_{\rm{s}}}}$, $\gamma_{\rm th}=2^{R_{\rm b}/W}-1$, $R_{\rm b}$ denotes the DT rate, $W$ is the transmission bandwidth, and $K_1(\cdot)$ is the first-order modified Bessel function of the second kind.
The successful transmission probability is given by ${P_{{\rm{succ}}}}=1 - {P_{{\rm{out}}}}$.
 For each IT, the EC of the BD is given by ${E_{\rm{T}}} = {P_{\rm{c}}}{T_{\rm{b}}}$, where ${P_{\rm{c}}}$ denotes the circuit power consumption.

\subsubsection{RT}
When the BD decodes a NACK feedback, it retransmits the previously generated packet stored at the BD. Since RT adopts the same full-reflection BackCom mode and suffers the same statistical channel as IT, the decoding SNR and OP during RT are identical to those of IT, which are respectively given by $\gamma \left( m \right)$ and ${P_{{\rm{out}}}}$ in \eqref{2A}. Moreover, the EC of RT is also the same as that of IT, which is given by ${E_{\rm{T}}}$.

\subsubsection{FD}
After each IT or RT, the R feeds back a one-bit ACK/NACK message to the BD, indicating whether the transmitted packet is successfully decoded. Under the outage-based decoding model, the feedback is determined as
\setlength{\abovedisplayskip}{2pt}
\setlength{\belowdisplayskip}{2pt}
\begin{align}\label{3A}
b\left( m \right)=
\begin{cases}
\mathrm{ACK}, & \text{if } \gamma (m) \ge {\gamma _{{\rm{th}}}},\\
\mathrm{NACK}, & \text{if } \gamma (m) < {\gamma _{{\rm{th}}}}.
\end{cases}
\end{align}
Considering the low-complexity architecture of the BD, a non-coherent energy detection based on orthogonal signaling is adopted for FD\cite{tse2005fundamentals}. Specifically, the ACK and NACK messages are respectively mapped into two orthogonal signals given by ${{\bf{x}}_{\rm{A}}(m)} = {\left[ {1 ,0} \right]^T}$ and ${{\bf{x}}_{\rm{N}}(m)} = {\left[ {0,1 } \right]^T}$.
 The received feedback signal at the BD is given by ${{\bf{y}}_{\rm{f}}(m)} = {\sqrt {P_{\rm f}}}g_1(m){{\bf{x}}_{\rm{f}}(m)} + {{\bf{w}}_{\rm{f}}(m)}$, where ${{\bf{x}}_{\rm{f}}(m)} \in \left\{ {{{\bf{x}}_{\rm{A}}(m)},{{\bf{x}}_{\rm{N}}}(m)} \right\}$, ${P_{\rm f}}$ is the feedback transmit power, and ${{\bf{w}}_{\rm{f}}(m)} \sim CN\left( {0,{\sigma _{\rm f}^2}{\bf{I}}} \right)$ denotes the AWGN. Due to the lack of instantaneous CSI of $g_1(m)$ at the BD, it performs non-coherent detection by comparing the received energies of the two orthogonal dimensions, i.e., $|y_1|^2\!\!\!\!
\underset{\mathrm{NACK}}
{\overset{\mathrm{ACK}}{\gtrless}}
\!\!\!\!|y_2|^2$, where ${{y_1}}$ and ${{y_2}}$ are the two elements of ${{\bf{y}}_{\rm{f}}(m)}$.

Let $\Omega_{\hat{\rm A}|\rm A}$ and $\Omega_{\hat{\rm N}|\rm N}$ denote the conditional FD events in which the BD correctly detects an ACK given an actual ACK and a NACK given an actual NACK, respectively. Similarly, $\Omega_{\hat{\rm N}|\rm A}$ and $\Omega_{\hat{\rm A}|\rm N}$ denote the conditional events of incorrectly detecting an ACK as a NACK and a NACK as an ACK, respectively. The corresponding conditional detection probabilities are denoted by $P_{\hat{\rm A}|\rm A}$, $P_{\hat{\rm N}|\rm N}$, $P_{\hat{\rm N}|\rm A}$, and $P_{\hat{\rm A}|\rm N}$, respectively. The FD error probabilities are given by\cite{tse2005fundamentals},
\setlength{\abovedisplayskip}{2pt}
\setlength{\belowdisplayskip}{2pt}
 \begin{align}\label{4A}
{P_{\left. {\widehat {\rm{A}}} \right|{\rm{N}}}} = {P_{\left. {\widehat {\rm{N}}} \right|{\rm{A}}}} = {\lambda _{g_1}}{{\sigma _{\rm f}^2} \mathord{\left/
 {\vphantom {{\sigma _{\rm f}^2} {\left( {{P_{\rm{f}}}\sigma _{\rm f}^2 + 2{\lambda _{g_1}}\sigma _{\rm f}^2} \right)}}} \right.
 \kern-\nulldelimiterspace} {\left( {{P_{\rm{f}}}+ 2{\lambda _{g_1}}\sigma _{\rm f}^2} \right)}}.
  \end{align}
and ${P_{\left. {\hat{\rm{A}}} \right|{\rm{A}}}} = 1 - {P_{\left. {\hat{\rm{N}}} \right|{\rm{A}}}}$, ${P_{\left. {\hat{\rm{N}}} \right|{\rm{N}}}} = 1 - {P_{\left. {\hat{\rm{A}}} \right|{\rm{N}}}}$.
Moreover, for the BD, the EC of FD in each transmission block is denoted by ${E_{{\rm{FD}}}}$, and the durations required for FD are assumed to be negligible compared with the duration of DT.


In summary, DG, IT, RT, and FD incur EC, denoted by ${E_{{\rm{DG}}}}$, ${E_{{\rm{T}}}}$, ${E_{{\rm{T}}}}$, and ${E_{{\rm{FD}}}}$, respectively, in each transmission block, while EH replenishes the stored energy by harvesting energy ${E_{\rm{H}}}$.
Both IT and RT are implemented through BackCom under a full-reflection mode, and thus the BD does not harvest energy during transmission processes.
In each transmission block, the BD performs only one of the three modes, i.e., IT, RT, and EH, where FD is performed only after IT or RT, and DG is performed only before IT, in the same transmission block.

\begin{figure*}[t]
  \centering
  \includegraphics[width=0.7\textwidth]{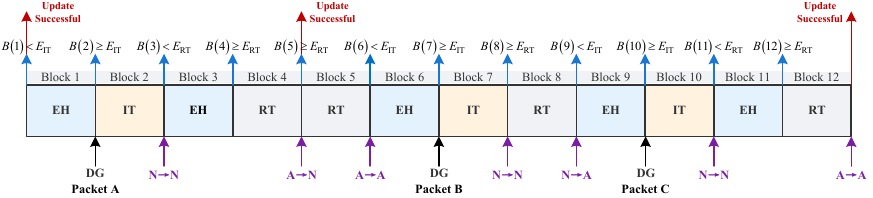}
  \vspace{-1em}
  \caption{Illustration of the BD's operation process under the proposed protocol.}
  \label{fig:transmission process}
  \vspace{-1em}
\end{figure*}

\vspace{-1em}
\section{Energy Management Protocol}\label{sec:Energy Management Protocol}
This section introduces the threshold-triggered EA-then-act energy management protocol. We first describe three operation modes of the BD in each transmission block under the proposed protocol, and then characterize the corresponding packet-buffer and stored-energy evolutions.

Before describing the operation modes, we first introduce two state variables, i.e., the stored energy of the BD $B\left( m \right)$ at the beginning of the $m$-th transmission block, and the packet buffer state $q\left( m \right) \in \left\{ {0,1} \right\}$. Specifically, $q\left( m \right) = 0$ indicates that no packet is buffered at the BD, while $q\left( m \right) = 1$ indicates that a previously generated packet remains in the BD buffer and is retained for RT. The packet buffer state is updated after IT or RT according to the feedback decoding results, while it remains unchanged after EH.


\vspace{-1em}
\subsection{Three Operation Modes of the BD }

There are three operation modes of the BD, namely \emph{``IT mode'', ``RT mode'', and ``EH mode''}, where only one mode is activated in each transmission block.

\subsubsection{IT mode}
The BD first performs DG through sensing operations, followed by IT and FD, while no energy is harvested in this mode. It operates in the IT mode in the $m$-th transmission block when $q\left( m \right) = 0$ and $B\left( m \right) \ge {E_{{\rm{IT}}}}$, where ${E_{{\rm{IT}}}} = {E_{{\rm{DG}}}} + {E_{\rm{T}}} + {E_{{\rm{FD}}}}$ denotes the energy threshold, equivalently the total EC required for the IT mode.

%
%

\subsubsection{RT mode}
The BD retransmits the previously generated
data packet without performing DG, followed by FD, and no energy is harvested during this mode. It operates in the RT mode in the $m$-th transmission block when $q\left( m \right) = 1$ and $B\left( m \right) \ge {E_{{\rm{RT}}}}$, where ${E_{{\rm{RT}}}} = {E_{\rm{T}}} + {E_{{\rm{FD}}}}$ denotes the energy threshold, equivalently the total EC required for the RT mode.

%
%

\subsubsection{EH mode}
 The BD harvests energy and accumulates
it across transmission blocks. It operates in this mode in the $m$-th transmission block when $\left\{ {q\left( m \right) = 0} \right\} \cap \left\{ {B\left( m \right) < {E_{{\rm{IT}}}}} \right\}$ or $\left\{ {q\left( m \right) = 1} \right\} \cap \left\{ {B\left( m \right) < {E_{{\rm{RT}}}}} \right\}$. The harvested energy of the BD in the $m$-th transmission block is ${E_{\rm{H}}}\left( m \right)$, and no EC.


\begin{definition}[Operation-Mode Rule for the BD]
The operation mode of the BD in the $m$-th transmission block is
\setlength{\abovedisplayskip}{2pt}
\setlength{\belowdisplayskip}{2pt}
\begin{align}\label{5A}
M\left( m \right)=
\begin{cases}
\mathrm{the\;IT\;mode}, & \rm{if }\; \Omega_{\rm IT},\\
\mathrm{the\;RT\;mode}, & \rm{if } \; \Omega_{\rm RT},\\
\mathrm{the\;EH\;mode}, & \rm{if } \; \Omega_{\rm EH},
\end{cases}
\end{align}
where events $\Omega_{\rm IT}$, $\Omega_{\rm RT}$, and $\Omega_{\rm EH}$ can be expressed as
\setlength{\abovedisplayskip}{2pt}
\setlength{\belowdisplayskip}{2pt}
\begin{equation}\label{6A}
\begin{aligned}
\Omega_{\rm IT}
&=
\{q\left(m\right)=0\}
\cap
\{B\left(m\right)\geq E_{\rm IT}\},
\\
\Omega_{\rm RT}
&=
\{q\left(m\right)=1\}
\cap
\{B\left(m\right)\geq E_{\rm RT}\},
\\
\Omega_{\rm EH}
&=
\left(
\{q\left(m\right)=0\}
\cap
\{B\left(m\right)<E_{\rm IT}\}
\right)\\
&\quad\cup
\left(
\{q\left(m\right)=1\}
\cap
\{B\left(m\right)<E_{\rm RT}\}
\right).
\end{aligned}
\end{equation}
\end{definition}


The above events are mutually exclusive and collectively exhaustive. Therefore, the operation mode of the BD is uniquely determined by the packet-buffer state and the stored-energy state at the beginning of each transmission block.
For clarity, Fig.~\ref{fig:transmission process} is provided to illustrate the BD's operation process under the proposed protocol.

\vspace{-1em}
\subsection{Evolution of the Packet-Buffer and Stored-Energy States}
The evolution of the packet-buffer state depends on the operating mode of the BD and the detected feedback results after IT or RT.
Specifically, there are the following cases:
\begin{itemize}
  \item When the BD operates in IT or RT mode in the $m$-th transmission block (${{\Omega _{{\rm{IT}}}} \cup {\Omega _{{\rm{RT}}}}}$) and the BD detects an ACK ($\Omega_{\hat{\rm A}}$), the packet is removed from the buffer since the BD considers that the packet has been successfully delivered. Therefore, the packet-buffer state at the beginning of the $\left(m+1\right)$-th transmission block is empty, i.e., $q\left(m + 1\right) = 0$.
  \item When the BD operates in IT or RT mode in the $m$-th transmission block (${{\Omega _{{\rm{IT}}}} \cup {\Omega _{{\rm{RT}}}}}$) and the BD detects a NACK ($\Omega_{\hat{\rm N}}$),  the packet is retained for subsequent RT since the BD considers that the packet has failed to deliver. Therefore, the packet-buffer state at the beginning of the $\left(m+1\right)$-th transmission block is non-empty, i.e., $q\left(m + 1\right) = 1$.
  \item When the BD operates in the EH mode in the $m$-th transmission block (${{\Omega _{{\rm{EH}}}}}$), the packet-buffer state remains unchanged, i.e., $q\left(m + 1\right) = q\left(m\right) $.
\end{itemize}
Accordingly, the packet-buffer state evolution is described as
\begin{align}\label{7A}
q\left(m\!\!+\!\!1\right)\!=\!
\begin{cases}
0,
&\!\!\text{if }
(\Omega_{\rm IT}
\cup
\Omega_{\rm RT})
\cap
\Omega_{\hat{\rm A}},
\\
1,
&\!\!\text{if }
(\Omega_{\rm IT}
\cup
\Omega_{\rm RT})
\cap
\Omega_{\hat{\rm N}},
\\
q\left(m\right),
&\!\!\text{if }
\Omega_{\rm EH}.
\end{cases}
\end{align}
Meanwhile, the evolution of the stored energy at the beginning of each transmission block is determined by the operating mode of the BD and constrained by the finite storage capacity, given by
\vspace{-1em}
\begin{align}\label{8A}
B(m\!+\!1)\!=\!
\begin{cases}
\min\{B(m)+E_{\rm H}(m),B_{\rm max}\},
& \text{if } \Omega_{\rm EH},\\
B(m)-E_{\rm IT},
& \text{if } \Omega_{\rm IT},\\
B(m)-E_{\rm RT},
& \text{if } \Omega_{\rm RT}.
\end{cases}
\end{align}

%
%
%

\section{Energy-State Evolution and Steady-State Distribution}\label{sec:Energy State Analysis and Steady-State Distribution}
This section characterizes the energy-state evolution of the BD and
derive its steady-state distribution at the beginning of each update cycle.
This analysis serves as the foundation for the subsequent AAoI analysis,
since the energy state determines the BD's operation mode, which further affects the update interval evolution.
To facilitate the analysis, the continuous
energy is discretized into a finite state space, based on which the energy evolution is modeled as a finite-state Markov chain.
Then, we characterize the energy-state transition matrices for a single
transmission attempt and further extend the analysis to a complete packet
delivery process and an entire update cycle. Finally, the steady-state energy
distribution is obtained.
\vspace{-1em}
\subsection{Discretization of Stored Energy State}
To characterize the evolution of the stored energy state of the BD, the continuous energy interval $[0,B_{\max}]$ is divided into $Q+1$ discrete energy levels, denoted by
$\mathcal{E}=\{e_0,e_1,\ldots,e_Q\}$. The discrete energy levels are defined as
$e_q=q\Delta, q=0,1,\ldots,Q$,
where $\Delta=B_{\max}/Q$ denotes the energy quantization interval. We note that there are some gaps between the discrete energy levels and the continuous stored energy, but such a gap becomes ignorable when $Q$ is sufficiently large. Here we adopt the downward quantization rule, such that the quantized energy state never overestimates the actual
stored energy and the energy-causality constraint is always satisfied. Specifically, the continuous stored energy $x$ is mapped to the
largest discrete energy level that does not exceed $x$, which is given by
$\xi(x)=e_{\min\{\lfloor x/\Delta\rfloor,Q\}}$,
where $x\in[e_q,e_{q+1})$ is quantized as $\xi(x)=e_q$.
In addition, we define $Q+1$ discrete energy states as
$\mathcal{S}=\{s_0,s_1,\ldots,s_Q\}$, where state $s_q$ corresponds to a
stored energy level $e_q$.
For the following analysis, we assume that the storage capacity is sufficient to support both transmission modes, i.e., ${B_{\max }} \ge {E_{{\rm{IT}}}} \ge {E_{{\rm{RT}}}}$, so that both IT and RT can eventually be activated through EA.

\vspace{-1em}
\subsection{Energy-State Transition Matrix for One Transmission Attempt}

Before analyzing the energy evolution over a complete update
cycle, we first derive the energy-state transition matrix associated
with one transmission attempt.

For notational simplicity, let $C\in\{E_{\rm IT},E_{\rm RT}\}$ denote a unified notation for the energy thresholds required for the IT and RT modes.
Suppose that the initial energy state is $s_a$, corresponding to stored energy $e_a$.
If $e_a<C$, the BD first performs EH over consecutive blocks until the stored energy reaches $C$, after which one transmission attempt with EC $C$ is performed.
Accordingly, the minimum number of EH blocks required to reach the threshold $C$ before the transmission attempt is defined as
\begin{align}\label{9A}
N_C(e_a)
\!=\!
\inf\!
\left\{\!\!
n\!\geq \!0\!:\!
\min\!
\left(\!\!
B_{\max},
e_a\!+\!\sum_{i=1}^{n}\!E_{\rm H}(i)
\!\right)
\!\!\geq \!\!C
\!\right\}.
\end{align}
Since $C\le B_{\max}$ and $E_H(i)>0$ almost surely, $N_C(e_a)$ is finite almost surely.

\begin{lemma}\label{lemma1}
Given the energy threshold $C$, the energy-state transition probability from the initial state $s_a$ to the final state $s_b$, through the required EH process and one transmission attempt with EC $C$, is given by
\begin{align}\label{10A}
P_{ab}^{(C)}
\!=\!
\begin{cases}
\!\mathbf{1}\{e_b=\xi(e_a-C)\},
&e_a\geq C,\\
\!e^{-\lambda_H e_b}
-e^{-\lambda_H e_{b+1}},
&e_a<C,\ 0\leq b<Q_C,\\
\!e^{-\lambda_H e_{Q_C}},
&e_a<C,\ b=Q_C,\\
\!0,
&e_a<C,\ b>Q_C,
\end{cases}
\end{align}
where ${\lambda _H} \!=\! {{{\lambda _f}} \mathord{\left/
 {\vphantom {{{\lambda _f}} {\left( {\eta P{T_b}} \right)}}} \right.
 \kern-\nulldelimiterspace} {\left( {\eta P_{\rm s}{T_b}} \right)}}$ is the rate parameter of the harvested energy $E_H(i)$, and $Q_C
=
\min
\left\{
\left\lfloor
\frac{B_{\max}-C}{\Delta}
\right\rfloor,
Q
\right\}$ denotes the largest reachable energy-state index after \!consuming \!energy $C$.
\end{lemma}

\begin{proof}
The proof is provided in Appendix A.
\end{proof}

Based on Lemma \ref{lemma1}, the energy-state transition matrices associated with one IT and one RT attempt are respectively given by
$\mathbf{P}_{\rm IT}=\mathbf{P}^{(C)}|_{C=E_{\rm IT}}$ and
$\mathbf{P}_{\rm RT}=\mathbf{P}^{(C)}|_{C=E_{\rm RT}}$, where ${{\bf{P}}^{\left( C \right)}} = {\left[ {P_{ab}^{(C)}} \right ] _{\left( {Q + 1} \right) \times \left( {Q + 1} \right)}}$.
An important property of \eqref{10A} is that, for all initial energy states satisfying $ e_a< C$, the post-transmission energy distribution is independent of $e_a$.
This follows from the memoryless property of the exponentially distributed harvested energy. Specifically, when the accumulated harvested energy first reaches or exceeds the energy deficit $C-e_a$, the overshoot $O = \sum\limits_{i = 1}^{{N_C}\left( {{e_a}} \right)} {{E_H}\left( i \right)}  - \left( {C - {e_a}} \right)$ follows an exponential distribution with rate $\lambda_H$, whose distribution does not depend on the value of $C-e_a$. After accounting for the finite storage capacity and the EC $C$, the residual energy after transmission is given by ${R_C} = \min \left\{ {{B_{\max }} - C,O} \right\}$. Therefore, the post-transmission energy distribution is identical for all initial energy states satisfying $ e_a< C$.

\vspace{-0.5em}
\subsection{Energy-State Transition Matrices for One Fresh Packet}\label{subsec:Energy-State Transition Matrices for One Fresh Packet}
The previous subsection characterizes the energy transition associated with a single transmission attempt. A fresh packet, however, may experience multiple transmission attempts because a failed transmission followed by a correctly detected NACK triggers RT.
We therefore extend the analysis from the transmission-attempt level to the packet level. In particular, we characterize the energy-state transition from the beginning of a fresh-packet transmission process until the packet is either successfully decoded at the R or discarded by the BD before successful decoding.

For notational convenience, let
$\gamma = P_{\rm out}P_{\hat {\rm{N}}|{\rm N}}$ denote the probability that a transmission attempt fails at the R and the resulting NACK is correctly detected at the BD, and $\delta = P_{\rm out}P_{\hat {\rm{A}}|{\rm N}}$ denote the probability that a transmission attempt fails at the R but the resulting NACK is incorrectly detected as an ACK at the BD.

\subsubsection{Successful Delivery or Packet Discard Starting from the RT Mode}
Starting from the RT mode, the packet can be successfully decoded at the R or discarded due to false ACK detection. In either case, the preceding $k$ RT attempts all fail with correctly detected NACKs, while the $(k+1)$-th RT attempt results in successful decoding or packet discard due to false ACK detection, respectively. The corresponding energy-state transition matrices are $\gamma^k P_{\rm succ}\mathbf P_{\rm RT}^{k+1}$ and $\gamma^k\delta\mathbf P_{\rm RT}^{k+1}$, respectively. Accordingly, the energy-state transition matrices $\mathbf S_r$ and $\mathbf D_r$ from the RT mode to eventual successful decoding and packet discard, respectively, are obtained as
\begin{equation}\label{11A}
\mathbf S_r
=
\sum_{k=0}^{\infty}
\gamma^k{P_{{\rm{succ}}}}\mathbf P_{\rm{RT}}^{k+1}
=
(\mathbf I_{Q+1}-\gamma\mathbf P_{\rm{RT}})^{-1}
{P_{{\rm{succ}}}}\mathbf P_{\rm{RT}},
\end{equation}
\begin{equation}\label{12A}
\mathbf D_r=
\sum_{k=0}^{\infty}
\gamma^k\delta\mathbf P_{\rm{RT}}^{k+1}
=
(\mathbf I_{Q+1}-\gamma\mathbf P_{\rm{RT}})^{-1}
\delta\mathbf P_{\rm{RT}} .
\end{equation}

\subsubsection{Fresh Packet Transition to Successful Delivery or Packet Discard}
Starting from the IT mode, a fresh packet may be successfully decoded at the R or discarded due to false ACK detection, either during IT or after subsequent RT attempts. The energy-state transition matrices for these two outcomes during IT are $P_{\rm succ}\mathbf P_{\rm IT}$ and $\delta\mathbf P_{\rm IT}$, respectively. If IT fails with a correctly detected NACK, the packet enters the RT mode, characterized by $\gamma\mathbf P_{\rm IT}$, where its subsequent outcomes are governed by $\mathbf S_r$ and $\mathbf D_r$. Therefore, the energy-state transition matrices $\mathbf S_p$ and $\mathbf D_p$ of a fresh packet to eventual successful decoding and packet discard are given by
\begin{equation}\label{13A}
\mathbf S_p
=
{P_{{\rm{succ}}}}\mathbf P_{\rm{IT}}
+\gamma\mathbf P_{\rm{IT}}\mathbf S_r,\quad
\mathbf D_p
=
\delta\mathbf P_{\rm{IT}}
+\gamma\mathbf P_{\rm{IT}}\mathbf D_r.
\end{equation}

\vspace{-1em}
\subsection{Energy-State Transition Matrix for a Complete Update Cycle}
We next characterize the energy-state evolution between two consecutive successful updates at the R.
Due to imperfect FD, a successful update at the R does not necessarily leave the BD buffer empty. Specifically, if the ACK associated with the previous successful update is incorrectly detected as a NACK, the already delivered packet remains in the BD buffer and is unnecessarily retransmitted at the beginning of the next update cycle.
Therefore, a cycle can start either with an empty buffer or with an already delivered packet remaining in the buffer.

\subsubsection{Update Cycle Starting from an Empty Buffer}
Suppose that the update cycle starts with an empty buffer. Before the next successful update occurs, one or more fresh packets may be discarded due to NACK-to-ACK feedback errors.

If exactly $k$ fresh packets are discarded before the $(k+1)$-th fresh packet is successfully decoded, the corresponding energy-state transition matrix is $\mathbf D_p^k\mathbf S_p$ for $k\ge 0$. Therefore, the energy-state transition matrix for an update cycle starting from an empty buffer can be obtained as
\begin{equation}\label{14A}
\begin{aligned}
\mathbf A_0
=
\sum_{k=0}^{\infty}\mathbf D_p^k\mathbf S_p
=
(\mathbf I_{Q+1}-\mathbf D_p)^{-1}\mathbf S_p .
\end{aligned}
\end{equation}

\subsubsection{Update Cycle Starting with an Already Delivered Packet}

For one RT attempt of the retained packet, let $\alpha = {P_{{\rm{succ}}}}P_{\hat {\rm{A}}|\rm{A}} + P_{\rm out}P_{\hat {\rm{A}}|\rm{N}}$ denote the probability that the BD detects an ACK and consequently clears the packet from its buffer, and $\beta = {P_{{\rm{succ}}}}P_{\hat {\rm{N}}|\rm{A}} + P_{\rm out}P_{\hat {\rm{N}}|\rm{N}}$ denote the probability that the BD detects a NACK and consequently retains the packet.
If the old packet is cleared after the $(k+1)$-th RT attempt, the preceding $k$ RT attempts all result in a NACK decision at the BD. The corresponding transition matrix is $\beta^k\alpha\mathbf P_{\rm{RT}}^{k+1}$ for $k\ge 0$. Accordingly, the energy-state transition matrix from an already delivered packet to empty-buffer state is given by
\begin{equation}\label{15A}
\mathbf C_o
=
\sum_{k=0}^{\infty}
\beta^k\alpha\mathbf P_{\rm{RT}}^{k+1}
=
(\mathbf I_{Q+1}-\beta\mathbf P_{\rm{RT}})^{-1}
\alpha\mathbf P_{\rm{RT}} .
\end{equation}
After the old packet is cleared, the system enters the empty-buffer process characterized by $\mathbf A_0$. Therefore, the energy-state transition matrix for an update cycle starting from an already delivered packet is given by
$\mathbf A_1
=
\mathbf C_o\mathbf A_0$.

\vspace{-1em}
\subsection{Steady-State Energy Distribution}
We construct an embedded Markov chain by observing the quantized stored-energy state immediately after each successful update at the R. Since the feedback generated by the R after a successful update is an ACK, the next update cycle starts with an empty buffer if the ACK is correctly detected at the BD with probability $P_{\hat {\rm A}|{\rm A}}$. Otherwise, if the ACK is incorrectly detected as a NACK, an already delivered packet remains in the BD buffer with probability $P_{\hat {\rm N}|{\rm A}}$. Hence, the one-step transition matrix of the embedded Markov chain is
\begin{equation}\label{16A}
\mathbf A
={P_{\hat {\rm A}|{\rm A}}}{\mathbf A_0}
+{P_{\hat {\rm N}|{\rm A}}}{\mathbf A_1}.
\end{equation}
Assuming that the resulting finite-state Markov chain is ergodic, its steady-state distribution is denoted by $\boldsymbol{\pi} = [\pi_0,\pi_1,\ldots,\pi_Q]$,
which satisfies
$\boldsymbol{\pi} = \boldsymbol{\pi}\mathbf A$,
together with the normalization condition
$\boldsymbol{\pi}\mathbf 1_{Q+1}=1$.


\vspace{-0.5em}
\section{AAoI Analysis}\label{sec:AAoI Analysis}
Based on the steady-state energy distribution derived in Section \ref{sec:Energy State Analysis and Steady-State Distribution}, we analyze the AAoI of the considered system. We first characterize the AoI evolution and derive a general AAoI expression. We then obtain the first- and second-order moments of the number of blocks required for one transmission attempt. Based on these results, the first and second moments of the inter-update interval are derived. Finally, the correlation between the system time of a successfully delivered packet and the subsequent inter-update interval is characterized, yielding to the final AAoI expression.
\begin{figure}
  \centering
  \includegraphics[width=0.6\linewidth]{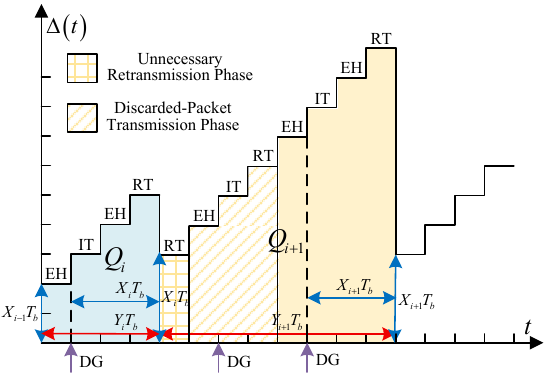}
  \vspace{-0.5em}
  \caption{AoI evaluation in Fig. \ref{fig:transmission process}.}\label{fig:AoI evaluation}
  \vspace{-1em}
\end{figure}

\vspace{-1em}
\subsection{AoI Evolution}
Let $r(t)$ denote the generation-block index of the most recently received status update at the end of transmission block $t$. The AoI at the R is given by $\Delta (t) = tT_{\rm{b}} - r(t)T_{\rm{b}}$. For a better understanding, the evolution of the instantaneous AoI corresponding to Fig.~\ref{fig:transmission process} is depicted as a staircase in Fig.~\ref{fig:AoI evaluation}.

Let $X_i$ denote the number of transmission blocks from the generation of the packet associated with the $i$-th successful update until its successful decoding at the R. Let $Y_i$ denote the number of transmission blocks between the $(i-1)$-th and $i$-th successful updates. Accordingly, the corresponding system time and inter-update interval are $X_iT_b$ and $Y_iT_b$, respectively.
Immediately after the $(i-1)$-th successful update, the AoI at the R decreases to $X_{i-1}T_b$.
It then increases by $T_b$ after each transmission block until the next successful update. Therefore, the AoI values over the $i$-th update cycle are $X_{i-1}T_b$, $(X_{i-1}+1)T_b$, $...$, $(X_{i-1}+{Y_i}-1)T_b$.
Accordingly, the cumulative AoI over the $i$-th update cycle can be obtained as
\begin{align}\label{17A}\notag
&{Q_i} = {T_b}\sum\limits_{k = 0}^{{Y_i} - 1} {\left( {{X_{i - 1}} + k} \right)}  = {T_b}\sum\limits_{k = 0}^{{Y_i} - 1} {{X_{i - 1}}}  + {T_b}\sum\limits_{k = 0}^{{Y_i} - 1} k \\
 &= {T_b}\!\left(\! \!{X_{i - 1}}{Y_i} \!+\! \frac{{{Y_i}\left( {{Y_i} \!-\! 1} \!\right)}}{2}\!\right) \!=\! {T_b}\!\left(\!\! {X_{i - 1}}{Y_i} \!+\! \frac{{Y_i^2}}{2}\! -\! \frac{{{Y_i}}}{2}\!\right).
\end{align}
The long-term AAoI is then given by
\begin{align}\label{18A}
\bar \Delta  = \frac{{\mathbb E\left[ {{Q_i}} \right]}}{{\mathbb E\left[ {{Y_i}} \right]}}= {T_b}\left(\frac{{\mathbb E\left[ {{X_{i - 1}}{Y_i}} \right]}}{{\mathbb E\left[ {{Y_i}} \right]}} + \frac{{\mathbb E\left[ {Y_i^2} \right]}}{{2 \mathbb E\left[ {{Y_i}} \right]}} - \frac{1}{2}\right).
\end{align}
It is worth noting that $X_{i-1}$ and $Y_i$ are generally correlated because the stored energy remaining after the $(i-1)$-th successful update affects the subsequent EA and transmission process. Therefore, $\mathbb E[X_{i-1}Y_i] \neq \mathbb E[X_{i-1}] \mathbb E[Y_i]$.
For notational simplicity, we use $X$ and $Y$ to denote generic stationary versions of the system time and inter-update interval in the following analysis. According to \eqref{18A}, the AAoI evaluation requires $\mathbb E[Y]$, $\mathbb E[Y^2]$ and $\mathbb E[XY]$, which will be evaluated in the following subsections.

\vspace{-1em}
\subsection{First- and Second-Order Moments of the Required Time for One Transmission Attempt}
For deriving the average update interval, we first characterize the statistical
moments of the blocks required for completing one transmission attempt.
Specifically, given an initial energy state $s_a$, corresponding to stored energy $e_a$, and an energy threshold $C\in\{E_{\rm IT},E_{\rm RT}\}$, the BD first performs EH for $N_C(e_a)$ blocks if ${e_a} < C$ until the stored energy reaches $C$, and then executes one transmission attempt. Accordingly, the total number of blocks required for completing one transmission attempt is expressed as $Z_C(e_a)=N_C(e_a)+1$.
The first- and second-order joint moments of the required blocks for one transmission attempt associated with the energy state transition $s_a\rightarrow s_b$ are respectively defined as
\begin{equation}\label{19A}
T_{ab}^{(C)}
\!\!=\!\!
\mathbb E\!
\left[\!
Z_C(e_a\!)\mathbf 1_{\{s_a\rightarrow s_b\}}
\!\right]\!,
U_{ab}^{(C)}
\!\!=\!\!
\mathbb E\!
\left[\!
Z_C^2(e_a\!)\mathbf 1_{\{s_a\rightarrow s_b\}}
\!\right].
\end{equation}

\begin{lemma}
\label{lemma2}
For a given initial energy level $e_a$ and an energy threshold $C$, the
first- and second-order joint moments of the number of blocks required for one
transmission attempt are respectively given by
\begin{equation}\label{20A}
T_{ab}^{(C)}
=
\begin{cases}
P_{ab}^{(C)}, & e_a \geq C, \\[1ex]
\left(\mu_a^{(C)}+2\right)P_{ab}^{(C)}, & e_a < C,
\end{cases}
\end{equation}
\begin{equation}\label{21A}
U_{ab}^{(C)}
=
\begin{cases}
P_{ab}^{(C)}, & e_a \geq C, \\[1ex]
\left[
\left(\mu_a^{(C)}\right)^2
+5\mu_a^{(C)}
+4
\right]P_{ab}^{(C)}, & e_a < C,
\end{cases}
\end{equation}
where
$\mu_a^{(C)}
=
\lambda_H(C-e_a), e_a<C$.
\end{lemma}

\begin{proof}
The proof is provided in Appendix B.
\end{proof}

Based on Lemma~\ref{lemma2}, the first- and second-order joint-moment matrices
associated with one transmission attempt in the IT and RT modes are respectively
denoted by
$\mathbf{T}_{\rm IT}=\mathbf{T}^{(C)}|_{C=E_{\rm IT}}$,
$\mathbf{U}_{\rm IT}=\mathbf{U}^{(C)}|_{C=E_{\rm IT}}$,
$\mathbf{T}_{\rm RT}=\mathbf{T}^{(C)}|_{C=E_{\rm RT}}$, and
$\mathbf{U}_{\rm RT}=\mathbf{U}^{(C)}|_{C=E_{\rm RT}}$.
The corresponding first-moment vectors are
$\boldsymbol{\tau}_{\rm IT}
=
\mathbf{T}_{\rm IT}\mathbf{1}_{Q+1}$ and
$\boldsymbol{\tau}_{\rm RT}
=
\mathbf{T}_{\rm RT}\mathbf{1}_{Q+1}$, and the second-moment vectors are
$\boldsymbol{\kappa}_{\rm IT}
=
\mathbf{U}_{\rm IT}\mathbf{1}_{Q+1}$ and
$\boldsymbol{\kappa}_{\rm RT}
=
\mathbf{U}_{\rm RT}\mathbf{1}_{Q+1}$,
where $\mathbf{1}_{Q+1}$ denotes the $(Q+1)\times1$ all-one vector.
Accordingly, their $a$-th elements represent the first and second moments of the numbers of blocks required for one IT and RT attempt, respectively, when the initial battery state is $s_a$.

\vspace{-1em}
\subsection{First Moment of the Inter-Update Interval $\mathbb E \left[Y\right]$ }
Due to imperfect FD, an update cycle may start either directly with a fresh-packet transmission process or with an already delivered packet retained at the BD due to a false NACK. Let $F\in\{0,1\}$ denote the update-cycle starting state, where $F\!\!=\!\!0$ denotes the former case and $F\!\!=\!\!1$ the latter.

\subsubsection{Average Inter-Update Interval for $F=0$}
The system directly enters the fresh-packet transmission process, which consists of two phases.

\begin{itemize}
    \item In phase $\mathsf{E}$, the BD buffer is empty, and the BD enters the IT mode. The corresponding energy-state transition matrix is $\mathbf P_{\rm IT}$, while the first- and second-order time-reward matrices are $\mathbf T_{\rm IT}$ and $\mathbf U_{\rm IT}$, respectively.

    \item In phase $\mathsf{R}$, the current packet has not yet been successfully decoded at the R and is retained at the BD after a correctly detected NACK. The BD enters the RT mode. The energy-state transition matrix is $\mathbf P_{\rm RT}$, and the first- and second-order time-reward matrices are $\mathbf T_{\rm RT}$ and $\mathbf U_{\rm RT}$, respectively.
\end{itemize}

Define
$\mathbf{g}_{0}
=
\begin{bmatrix}
\mathbf{g}_{\mathsf{E}}\\
\mathbf{g}_{\mathsf{R}}
\end{bmatrix}$,
where the $a$-th elements of $\mathbf{g}_{\mathsf{E}}$ and
$\mathbf{g}_{\mathsf{R}}$ denote the expected remaining numbers of blocks
until the next successful update when the system starts from battery state
$s_a$ in phases $\mathsf{E}$ and $\mathsf{R}$, respectively.
Starting from phase $\mathsf E$, the BD first completes one IT attempt. If the packet is successfully decoded at the R, the current inter-update interval terminates. Otherwise, if $\Omega_{\hat{\rm A}|{\rm N}}$ occurs, the packet is discarded and the system returns to phase $\mathsf E$, corresponding to the transition $\delta\mathbf P_{\rm IT}$; if $\Omega_{\hat{\rm N}|{\rm N}}$ occurs, the packet is retained and the system enters phase $\mathsf R$, corresponding to $\gamma\mathbf P_{\rm IT}$. Therefore,
$\mathbf g_E = \boldsymbol{\tau}_{\rm IT} + \delta\mathbf P_{\rm IT}\mathbf g_E + \gamma\mathbf P_{\rm IT}\mathbf g_R$.
%
%
Similarly,
$\mathbf g_R = \boldsymbol{\tau}_{\rm RT} + \delta\mathbf P_{\rm RT}\mathbf g_E + \gamma\mathbf P_{\rm RT}\mathbf g_R$.


For compact representation, define
$\boldsymbol{\tau}_{0}
=
\begin{bmatrix}
\boldsymbol{\tau}_{\rm IT}\\
\boldsymbol{\tau}_{\rm RT}
\end{bmatrix}$,
and
$\mathbf{Q}_{0}
=
\begin{bmatrix}
\delta\mathbf{P}_{\rm IT}
&
\gamma\mathbf{P}_{\rm IT}
\\
\delta\mathbf{P}_{\rm RT}
&
\gamma\mathbf{P}_{\rm RT}
\end{bmatrix}$.
Then, $\mathbf g_E$ and $\mathbf g_R$ can be jointly expressed as
$\mathbf{g}_{0}
=
\boldsymbol{\tau}_{0}
+
\mathbf{Q}_{0}\mathbf{g}_{0}$,
which yields
\begin{align}\label{22A}
\mathbf{g}_{0}
=
\left(
\mathbf{I}_{2(Q+1)}
-
\mathbf{Q}_{0}
\right)^{-1}
\boldsymbol{\tau}_{0}.
\end{align}
Note that $\mathbf{Q}_{0}$ only contains the non-absorbing transitions before
the next successful update. The probabilities of the events
${\Omega _{\hat{\rm{A}}\left| {\rm{A}} \right.}}$ and ${\Omega _{\hat{\rm{N}}\left| {\rm{A}} \right.}}$ are excluded because both correspond to successful decoding at the R and thus terminate the current inter-update interval. Since a fresh-packet process always starts from phase $\mathsf{E}$, we define the selection matrix
$\mathbf{E}_{0}
=
\begin{bmatrix}
\mathbf{I}_{Q+1}
&
\mathbf{0}_{(Q+1)\times(Q+1)}
\end{bmatrix}$.
Accordingly, the conditional first-moment vector of the inter-update interval for a fresh-packet process is given by
\begin{align}\label{23A}
\mathbf{y}_{0}
=
\mathbf{E}_{0}\mathbf{g}_{0},
(\mathbf{y}_{0})_a
=
\mathbb{E}
\left[
Y\mid S=s_a,F=0
\right].
\end{align}
\subsubsection{Average Inter-Update Interval for $F=1$}
An already delivered packet remains at the BD due to a false NACK associated with the previous successful update. Since this packet has already been successfully decoded at the R, its subsequent RTs are unnecessary and do not terminate the current inter-update interval. The BD therefore continues retransmitting the retained packet until an ACK is detected and the packet is cleared from the buffer, after which the system enters the fresh-packet transmission process.
For each RT of the retained packet, define
$
\alpha
\!=\!
P_{\rm succ}P_{\hat {\rm A}| {\rm A}}
\!+\!
P_{\rm out}P_{\hat {\rm A}|{\rm N}}$ and
$\beta
\!=\!
P_{\rm succ}P_{\hat {\rm N}|{\rm A}}
\!+\!
P_{\rm out}P_{\hat {\rm N}|{\rm N}}$,
where $\alpha$ denotes the probability that the BD detects an ACK and hence
clears the retained packet, whereas $\beta$ denotes the probability that the BD
detects a NACK and consequently retains it for another RT.

Let $\mathbf y_1$ denote the conditional first-moment vector of the inter-update interval with an already delivered packet, whose $a$-th element is $(\mathbf{y}_{1})_a
=
\mathbb{E}
\left[
Y
\mid
S=s_a,F=1
\right]$.
The BD first performs one RT attempt for
the retained packet, and the corresponding mean number of required blocks is characterized by
$\boldsymbol{\tau}_{\rm RT}$. After the RT attempt, if the BD detects a
NACK, which occurs with probability $\beta$, the packet remains in the buffer and the same RT process continues, with the remaining inter-update interval characterized by $\mathbf y_1$.
If the BD detects an ACK with probability $\alpha$, the packet is cleared and the system enters the fresh-packet transmission process characterized by $\mathbf y_0$.
Accordingly,
%
$
\mathbf{y}_{1}
=
\boldsymbol{\tau}_{\rm RT}
+
\beta\mathbf{P}_{\rm RT}\mathbf{y}_{1}
+
\alpha\mathbf{P}_{\rm RT}\mathbf{y}_{0}$,
yielding
\begin{align}\label{24A}
\mathbf{y}_{1}
=
\left(
\mathbf{I}_{Q+1}
-
\beta\mathbf{P}_{\rm RT}
\right)^{-1}
\left(
\boldsymbol{\tau}_{\rm RT}
+
\alpha\mathbf{P}_{\rm RT}\mathbf{y}_{0}
\right).
\end{align}

\subsubsection{Average Inter-Update Interval}
The conditional first-moment vector of the
inter-update interval is given by
$\mathbf{y}
=
P_{\hat {\rm A}| {\rm A}}\mathbf{y}_{0}
+
P_{\hat {\rm N}| {\rm A}}\mathbf{y}_{1}$,
where $\mathbf y_0$ and $\mathbf y_1$ correspond to the cases $F=0$ and $F=1$, respectively. Averaging over the steady-state distribution $\boldsymbol{\pi}$, the first moment of the inter-update interval is obtained as
$\mathbb{E}[Y]
=
\boldsymbol{\pi}\mathbf{y}$.

\vspace{-1em}
\subsection{Second Moment of Inter-Update Interval $\mathbb E \left[Y^2\right]$ }

\subsubsection{Second Moment of the Inter-Update Interval for $F=0$}
Let $Z$ denote the number of blocks required for the current transmission attempt and let $Y'$ denote the remaining number of blocks after the current attempt if the update cycle has not yet terminated. Then, $Y=Z+Y'$,
and hence $Y^2=Z^2+2ZY'+(Y')^2$.


Define
$\mathbf{h}_{0}
=
\begin{bmatrix}
\mathbf{h}_{\mathsf{E}}\\
\mathbf{h}_{\mathsf{R}}
\end{bmatrix}$,
where the $a$-th elements of $\mathbf{h}_{\mathsf{E}}$ and
$\mathbf{h}_{\mathsf{R}}$ denote the conditional second moments of the
remaining number of blocks until the next successful update when the system
starts from battery state $s_a$ in phases $\mathsf{E}$ and $\mathsf{R}$,
respectively.
Starting from phase $\mathsf E$, the BD first completes one IT attempt. According to
$Y^2=Z^2+2ZY'+(Y')^2$, the resulting second moment consists of the second moment of the duration of the current IT attempt, the cross term between the current-attempt duration and the subsequent remaining time, and the second moment of the subsequent remaining time.
Therefore,
$\mathbf h_E =\boldsymbol{\kappa}_{\rm IT} + 2\delta\mathbf T_{\rm IT}\mathbf g_E + 2\gamma\mathbf T_{\rm IT}\mathbf g_R+ \delta\mathbf P_{\rm IT}\mathbf h_E + \gamma\mathbf P_{\rm IT}\mathbf h_R$.
Similarly,
$\mathbf h_R =\boldsymbol{\kappa}_{\rm RT} + 2\delta\mathbf T_{\rm RT}\mathbf g_E +2\gamma\mathbf T_{\rm RT}\mathbf g_R+\delta\mathbf P_{\rm RT}\mathbf h_E +\gamma\mathbf P_{\rm RT}\mathbf h_R$.

For compact representation, define
$\boldsymbol{\kappa}_{0}
=
\begin{bmatrix}
\boldsymbol{\kappa}_{\rm IT}\\
\boldsymbol{\kappa}_{\rm RT}
\end{bmatrix}$,
and
$\mathbf{W}_{0}
=
\begin{bmatrix}
\delta\mathbf{T}_{\rm IT}
&
\gamma\mathbf{T}_{\rm IT}
\\
\delta\mathbf{T}_{\rm RT}
&
\gamma\mathbf{T}_{\rm RT}
\end{bmatrix}$.
Using the transient transition matrix $\mathbf{Q}_{0}$ defined previously,
$\mathbf h_E$ and $\mathbf h_R$ can be jointly expressed as
$\mathbf{h}_{0}
=
\boldsymbol{\kappa}_{0}
+
2\mathbf{W}_{0}\mathbf{g}_{0}
+
\mathbf{Q}_{0}\mathbf{h}_{0}$, which gives
\begin{align}\label{25A}
\mathbf{h}_{0}
=
\left(
\mathbf{I}_{2(Q+1)}
-
\mathbf{Q}_{0}
\right)^{-1}
\left(
\boldsymbol{\kappa}_{0}
+
2\mathbf{W}_{0}\mathbf{g}_{0}
\right).
\end{align}
Since the case $F=0$ always starts from phase $\mathsf E$, the corresponding conditional second-moment vector of the inter-update interval is obtained as
\begin{align}\label{26A}
\mathbf{m}_{0}
=
\mathbf{E}_{0}\mathbf{h}_{0},
(\mathbf{m}_{0})_a
=
\mathbb{E}
\left[
Y^2
\mid
S=s_a,F=0
\right].
\end{align}

\subsubsection{Second Moment of the Inter-Update Interval for $F=1$}
Let $\mathbf m_1$ denote the corresponding conditional second-moment vector of the inter-update interval, whose $a$-th element is
$(\mathbf{m}_{1})_a
=
\mathbb{E}
\left[
Y^2
\mid
S=s_a,F=1
\right]$.
The BD first performs one RT attempt for the retained packet. If the BD detects a NACK with probability $\beta$, the packet remains in the buffer and the subsequent first- and second-moment vectors are characterized by $\mathbf y_1$ and $\mathbf m_1$, respectively. If the BD detects an ACK with probability $\alpha$, the retained packet is cleared and the system enters the fresh-packet transmission process, characterized by $\mathbf y_0$ and $\mathbf m_0$. Accordingly,
%
$\mathbf{m}_{1}
\!=\!
\boldsymbol{\kappa}_{\rm RT}
\!+\!
2\beta\mathbf{T}_{\rm RT}\mathbf{y}_{1}
\!+\!
2\alpha\mathbf{T}_{\rm RT}\mathbf{y}_{0}
\!+\!
\beta\mathbf{P}_{\rm RT}\mathbf{m}_{1}
\!+\!
\alpha\mathbf{P}_{\rm RT}\mathbf{m}_{0}$.
Solving the above recursion gives
\begin{align}
\label{27A}
\mathbf{m}_{1}
&=
\left(
\mathbf{I}_{Q+1}
-
\beta\mathbf{P}_{\rm RT}
\right)^{-1}
\Big(
\boldsymbol{\kappa}_{\rm RT}
+
2\beta\mathbf{T}_{\rm RT}\mathbf{y}_{1}
\nonumber\\
&\qquad\qquad\qquad
+
2\alpha\mathbf{T}_{\rm RT}\mathbf{y}_{0}
+
\alpha\mathbf{P}_{\rm RT}\mathbf{m}_{0}
\Big).
\end{align}

\subsubsection{Second Moment of the Inter-Update Interval}
The conditional second-moment vector of the inter-update interval is given by
$
\mathbf m
=
P_{\hat{\rm A}|{\rm A}}\mathbf m_0
+
P_{\hat{\rm N}|{\rm A}}\mathbf m_1.
$
Averaging over the steady-state energy distribution $\boldsymbol{\pi}$ gives
$
\mathbb E[Y^2]
=
\boldsymbol{\pi}\mathbf m.
$

\vspace{-1em}
\subsection{Cross Moment between System Time and Inter-Update Interval $\mathbb E \left[XY\right]$}

Considering the correlation between $X$ and $Y$, we first derive the joint reward matrix of the effective transmission time $X$ and the battery state associated with a
successful packet delivery.
Based on this
result, we further account for possible packet discards before the successful
packet and obtain the overall $X$-reward matrix between two consecutive
successful updates. Finally, combining this matrix with the conditional
first-moment vector of the subsequent inter-update interval derived above
yields the cross moment $\mathbb{E}[XY]$.
\subsubsection{Transmission-Time Reward from the RT Phase Until Successful Delivery}
Before deriving the transmission-time reward, we first characterize the
battery-state transition associated with the eventual successful delivery of
a packet starting from the RT phase as $\mathbf{H}_{\rm RT}$. By first-step decomposition,
$\mathbf{H}_{\rm RT}
=
P_{\rm succ}\mathbf{P}_{\rm RT}
+
\gamma\mathbf{P}_{\rm RT}\mathbf{H}_{\rm RT}$,
which gives
$\mathbf{H}_{\rm RT}
=
\left(
\mathbf{I}_{Q+1}
-
\gamma\mathbf{P}_{\rm RT}
\right)^{-1}
P_{\rm succ}\mathbf{P}_{\rm RT}$.

Let $\mathbf{L}_{\rm RT}$ denote the accumulated transmission-time reward of a fresh
packet that has entered the RT phase, whose $(a,b)$-th element is
$\left(\mathbf{L}_{\rm RT}\right)_{ab}
=
\mathbb{E}
\left[
X_{\rm RT}
\mathbf{1}_{\{S_{\rm succ}=s_b\}}
\mid
S=s_a,\mathsf{R}
\right]$,
where $X_{\rm RT}$ denotes the accumulated transmission time from the beginning
of the current RT attempt until the considered packet is successfully decoded
at the R, and $S_{\rm succ}$ denotes the battery state immediately after this
successful delivery.

Starting from the RT phase, the BD first performs one RT attempt for the considered packet. If the packet is successfully decoded at the R, the current RT duration contributes to $X_{\rm RT}$, yielding the term $P_{\rm succ}\mathbf T_{\rm RT}$. If the RT attempt fails and the resulting NACK is correctly detected at the BD with probability $\gamma$, the packet is retained and the RT process continues. In this case, the current RT duration contributes to $X_{\rm RT}$ provided that the packet is eventually successfully delivered, resulting in the term $\gamma\mathbf T_{\rm RT}\mathbf H_{\rm RT}$, while the subsequent RT process contributes its own accumulated transmission-time reward, characterized by $\gamma\mathbf P_{\rm RT}\mathbf L_{\rm RT}$. In contrast, if the failed RT is followed by a false ACK, the packet is discarded and does not contribute to $\mathbf L_{\rm RT}$. Hence,
$\mathbf{L}_{\rm RT}
\!\!=\!\!
P_{\rm succ}\!\mathbf{T}_{\rm RT}
\!+\!
\gamma\mathbf{T}_{\rm RT}\!\mathbf{H}_{\rm RT}
\!+\!
\gamma\mathbf{P}_{\rm RT}\!\mathbf{L}_{\rm RT}$, yielding
\begin{align}\label{28A}
\mathbf{L}_{\rm RT}
=
\left(
\mathbf{I}_{Q+1}
-
\gamma\mathbf{P}_{\rm RT}
\right)^{-1}
\left(
P_{\rm succ}\mathbf{T}_{\rm RT}
+
\gamma\mathbf{T}_{\rm RT}\mathbf{H}_{\rm RT}
\right).
\end{align}
\subsubsection{$X$-Reward Matrix of a Successfully Delivered Fresh Packet}
Let $\mathbf J$ denote the corresponding reward matrix, whose $(a,b)$-th element is
$J_{ab}
=
\mathbb{E}
\left[
X
\mathbf{1}_{\{S_{\rm succ}=s_b\}}
\mid
S=s_a
\right]$, where $X$ denotes the effective transmission time of the considered fresh packet, measured from the beginning of its IT attempt until its successful delivery at the R.

Starting from the IT phase, the BD first performs one IT attempt. If the packet is successfully decoded at the R, the corresponding contribution is $P_{\rm succ}\mathbf P_{\rm IT}$. If the IT fails and the resulting NACK is correctly detected at the BD with probability $\gamma$, the packet is retained and enters the RT phase. In this case, the IT block contributes $\gamma\mathbf P_{\rm IT}\mathbf H_{\rm RT}$, while the subsequent RT process contributes $\gamma\mathbf P_{\rm IT}\mathbf L_{\rm RT}$. Therefore, the $X$-reward matrix of a successfully delivered fresh packet is
\begin{equation}\label{29A}
\mathbf{J}
=
P_{\rm succ}\mathbf{P}_{\rm IT}
+
\gamma\mathbf{P}_{\rm IT}\mathbf{H}_{\rm RT}
+
\gamma\mathbf{P}_{\rm IT}\mathbf{L}_{\rm RT}.
\end{equation}
Here $\mathbf{P}_{\rm IT}$, rather than
$\mathbf{T}_{\rm IT}$, is used in \eqref{29A}. This is because
$\mathbf{T}_{\rm IT}$ accounts for the entire number of blocks required to
reach the IT energy threshold and complete the IT attempt, including the EH
waiting time before the fresh packet starts its IT.
Such pre-transmission waiting time does not belong to the effective
transmission time $X$. In contrast, once the IT starts, the
IT attempt itself contributes exactly one block to $X$.

\subsubsection{Overall $X$-Reward Matrix}
Before the successfully delivered fresh packet, one or more fresh packets may be discarded due to erroneous feedback detection. Although the transmission durations of these discarded packets do not contribute to the effective transmission time $X$ of the ultimately successful packet, they affect the energy state from which the successful packet transmission starts. Recall that $\mathbf D_p$ denotes the energy-state transition matrix of a fresh packet from its IT to its eventual discard. Therefore, for the case $F=0$, if exactly $k$ fresh packets are discarded before the next successfully delivered packet, the corresponding $X$-reward matrix is $\mathbf D_p^k\mathbf J$, $k\ge 0$, where $\mathbf{J}$ denotes the $X$-reward matrix of a fresh packet that is eventually successfully delivered.
Therefore, the overall $X$-reward matrix for an update cycle starting from a fresh-
packet process is given by
%
%
\begin{align}\label{30A}
\mathbf{K}_{0}
=
\sum_{k=0}^{\infty}
\mathbf{D}_{p}^{k}\mathbf{J}
=
\left(
\mathbf{I}_{Q+1}
-
\mathbf{D}_{p}
\right)^{-1}
\mathbf{J}.
\end{align}

For the case $F=1$, an already delivered packet remains at the BD and must be cleared before the fresh-packet transmission process begins. Since the RT of this retained packet do not contribute to the effective transmission time $X$ of the next successfully delivered fresh packet, they only affect the energy state at the beginning of the subsequent fresh-packet process. Recall that $\mathbf C_o$ denotes the energy-state transition matrix associated with clearing the retained packet. Hence, the overall $X$-reward matrix for an update cycle starting with an already delivered packet is given by
$\mathbf{K}_{1}
=
\mathbf{C}_{o}\mathbf{K}_{0}$.

%

Accordingly, the overall $X$-reward matrix is given by
\begin{equation}\label{31A}
\mathbf{K}
\!=\!
P_{\hat {\rm A}| {\rm A}}\mathbf{K}_{0}
\!+\!
P_{\hat {\rm N}| {\rm A}}\mathbf{K}_{1}
\!=\!
\left(
P_{\hat {\rm A}| {\rm A}}\mathbf{I}_{Q+1}
\!+\!
P_{\hat {\rm N}| {\rm A}}\mathbf{C}_{o}
\right)
\mathbf{K}_{0}.
\end{equation}

\subsubsection{Cross Moment of $X$ and $Y$}
Recall that the $(a,b)$-th element of $\mathbf K$ represents the joint $X$-reward associated with the energy-state transition from $s_a$ at the beginning of an update cycle to $s_b$ at the corresponding successful update, i.e.,
$K_{ab}
=
\mathbb{E}
\left[
X_i
\mathbf{1}_{\{S_i=s_b\}}
\mid
S_{i-1}=s_a
\right]$.
Under the steady-state energy distribution
$\boldsymbol{\pi}$,
$(\boldsymbol{\pi}\mathbf{K})_b
=
\mathbb{E}
\left[
X_i
\mathbf{1}_{\{S_i=s_b\}}
\right]$.
Moreover, from the first-moment analysis of the inter-update interval, the $b$-th element of $\mathbf{y}$ is given by
$(\mathbf{y})_b
=
\mathbb{E}
\left[
Y_{i+1}
\mid
S_i=s_b
\right]$.

Since the block-wise EH, channel, and feedback processes are independent across update cycles, conditioned on the energy state $S_i$ at the $i$-th successful update, the past process determining $X_i$ and the subsequent process determining $Y_{i+1}$ are conditionally independent. Therefore,
\begin{align}\notag
\mathbb{E}
\left[
X_iY_{i+1}
\right]
&=
\sum_{b=0}^{Q}
\mathbb{E}
\left[
X_i
\mathbf{1}_{\{S_i=s_b\}}
\right]
\mathbb{E}
\left[
Y_{i+1}
\mid
S_i=s_b
\right]
\\\label{32A}
&=
\sum_{b=0}^{Q}
(\boldsymbol{\pi}\mathbf{K})_b
(\mathbf{y})_b
=
\boldsymbol{\pi}\mathbf{K}\mathbf{y}.
\end{align}
By stationarity of the update process,
$\mathbb{E}[X_iY_{i+1}]
=
\mathbb{E}[X_{i-1}Y_i]=
\boldsymbol{\pi}\mathbf{K}\mathbf{y}$.

Based on the above derivations, the first and second moments of the
inter-update interval are respectively given by
$\mathbb{E}[Y]
=
\boldsymbol{\pi}\mathbf{y}$ and
$\mathbb{E}[Y^2]
=
\boldsymbol{\pi}\mathbf{m}$,
and the cross moment between the effective transmission time of the previously
successfully delivered packet and the subsequent inter-update interval is
$\mathbb{E}[X_{i-1}Y_i]
=
\boldsymbol{\pi}\mathbf{K}\mathbf{y}$.
Accordingly, the AAoI is obtained as
\begin{align}
\bar{\Delta}
=
{T_b}\left(\frac{
\boldsymbol{\pi}\mathbf{K}\mathbf{y}
}{
\boldsymbol{\pi}\mathbf{y}
}
+
\frac{
\boldsymbol{\pi}\mathbf{m}
}{
2\boldsymbol{\pi}\mathbf{y}
}
-
\frac{1}{2}\right).
\label{33A}
\end{align}

\section{PLR Analysis}\label{sec:PLR Analysis}
This section characterizes the reliability by the PLR, defined as the probability that a newly generated status-update packet is discarded before being successfully delivered to the R.

For a packet that has not yet been successfully delivered, a failed transmission followed by a correctly detected NACK occurs with probability $\gamma$, in which case the packet is retained for RT. In contrast, a failed transmission followed by a NACK-to-ACK error occurs with probability $\delta$, resulting in the premature removal of the undelivered packet. Accordingly, the probability that a packet is lost after $k$ correctly detected NACKs is ${\gamma}^k{\delta}$. Accordingly, the PLR is obtained as
\begin{align}
{P_{\rm{L}}} = \sum\limits_{k = 0}^\infty  {{\gamma ^k}} \delta  = {\delta  \mathord{\left/
 {\vphantom {\delta  {\left( {1 - \gamma } \right)}}} \right.
 \kern-\nulldelimiterspace} {\left( {1 - \gamma } \right)}}.
\label{34A}
\end{align}
The PLR is determined by the transmission reliability and the NACK-to-ACK detection error probability.
Besides, the PLR does not explicitly depend on the EA, since insufficient energy only postpones the transmission attempt rather than discarding the packet under the proposed protocol.

\section{Simulation Results}\label{sec:Simulation Results}
In this section, we present simulation results to evaluate the performance of the considered BackCom system. Each Monte Carlo simulation is performed over $10^{6}$ successful status update cycles. Following the channel model assumed in Section \ref{sec:System Model}, the large-scale path losses of the S$\to$BD and BD$\to$R links for DT, as well as that of the R$\to$BD link for FD, are calculated as follows \cite{11482890,8907447}:
\begin{align}
{L_f} \!= \!\frac{{\lambda _c^2{G_{\rm{S}}}{G_{{\rm{BD}}}}}}{{{{\left( {4\pi } \right)}^2}d_f^{{v_f}}}},{L_g} \!= \! \frac{{\lambda _c^2{G_{{\rm{BD}}}}{G_{\rm{R}}}}}{{{{\left( {4\pi } \right)}^2}d_g^{{v_g}}}},{L_{g_1}} \!=\! \frac{{\lambda _{c1}^2{G_{\rm{R}}}{G_{\rm{BD}}}}}{{{{\left( {4\pi } \right)}^2}d_g^{{v_g}}}},
\end{align}
where ${{\lambda _c}}$ and ${{\lambda _{c1}}}$ denote the carrier wavelengths for DT and FD, respectively; ${{G_{\rm{S}}}}$, ${{G_{\rm{BD}}}}$, and ${{G_{\rm{R}}}}$ are the antenna gains for the S, the BD, and the R, respectively; ${d_f}$ and ${d_g}$ are the distances of the likes of ${\rm{S}} \to {\rm{BD}}$ and ${\rm{BD}} \to {\rm{R}}$, respectively; and ${{v_i}}$ with $i = \left\{ {f,g} \right\}$ is the pass loss exponent of channel $i$.
Following \cite{3gpp38769}, we set ${{\lambda _c}}=0.33$ m corresponding to 900 MHz,  ${{\lambda _{c1}}}=0.317$ m corresponding to 945 MHz, ${{G_{\rm{S}}}}=6$ dBi, ${{G_{\rm{BD}}}}=0$ dBi, ${{G_{\rm{R}}}}=6$ dBi, ${{v_i}}=3.5$, and the 2-D Cartesian
coordinates of the S, the BD, and the R are set to $\left( {-3,4} \right)$ m, $\left( {0,0} \right)$ m and $\left( {12,9} \right)$ m, respectively.
The RF source transmit power is set to ${P_{\rm{s}}} = 24$ dBm.
For DT, the channel bandwidth and DT rate are set to $W = 15$ kHz and ${R_{\rm{b}}} = 5$ kbps, respectively, and the equivalent noise power at R is $\sigma _t^2 =  - 110$ dBm. For FD, the feedback power at R and the equivalent noise power at BD are set to ${P_{\rm{f}}} = 13$ dBm and $\sigma _{\rm{f}}^2 =  - 70$ dBm.
Moreover, the maximum energy-storage capacity is set to ${B_{\max }} = 100$ nJ, and the duration of each transmission block is ${T_{\rm{b}}} = 20$ ms. The energy conversion efficiency is $\eta  = 0.3$, while the circuit power consumption of the BD is ${P_{\rm{c}}} = 1$ uW.
The ECs for DG and FD are set to ${E_{{\rm{DG}}}} = 5$ nJ and ${E_{{\rm{FD}}}} = 2.5$ nJ, respectively. Unless otherwise specified, all parameters are kept unchanged in simulations.

\begin{figure}
  \centering
  \includegraphics[width=0.65\linewidth]{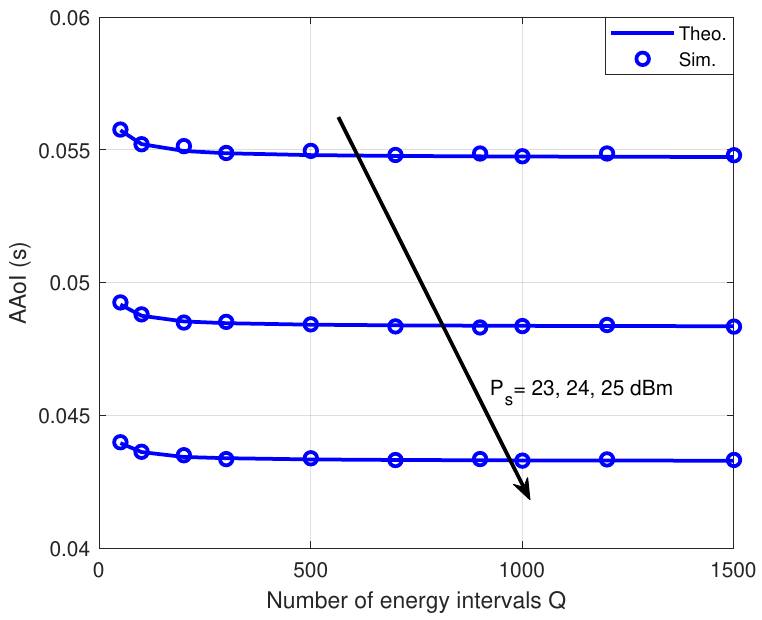}
    \vspace{-1em}
  \caption{Convergence of AAoI versus the number of energy intervals $Q$.}\label{Fig3}
  \vspace{-1em}
\end{figure}
\begin{figure}
  \centering
  \includegraphics[width=0.67\linewidth]{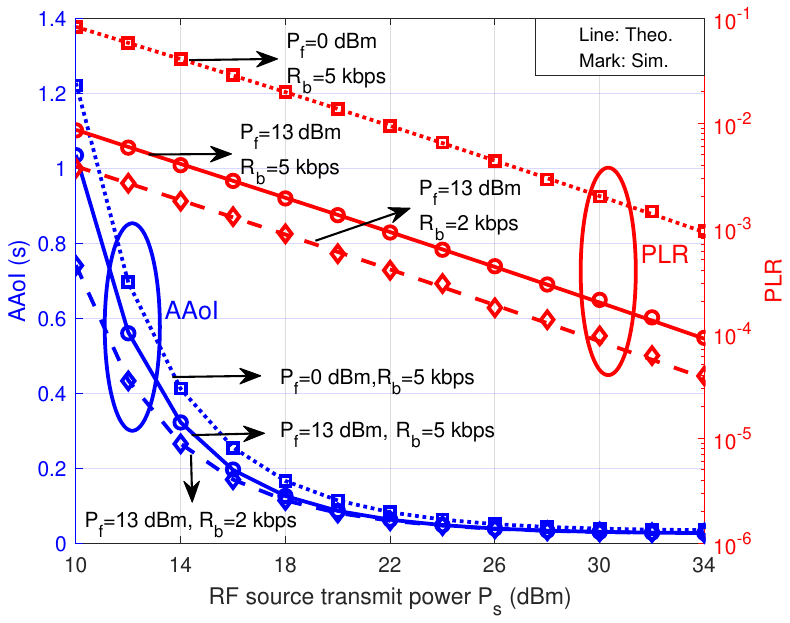}
    \vspace{-1em}
  \caption{Validation of theoretical AAoI and PLR.}\label{Fig4}
    \vspace{-1em}
\end{figure}

Fig. \ref{Fig3} illustrates the convergence of AAoI with respect to the number of energy quantization intervals $Q$. As $Q$ increases, the theoretical AAoI gradually converges and closely matches the Monte Carlo simulation results for all considered values of $P_{\rm s}$, validating the accuracy of the derived AAoI expression and indicating that the error introduced by energy quantization becomes negligible for sufficiently large $Q$. Accordingly, $Q=1000$ is adopted in the subsequent simulations.

Fig. \ref{Fig4} compares the theoretical and Monte Carlo simulation results for AAoI and PLR under different feedback transmit powers $P_{\rm f}$ and DT rates $R_{\rm b}$. The close agreement between theoretical and simulation results validates the derived expressions. As $P_{\rm s}$ increases, both AAoI and PLR decrease. This is because higher $P_{\rm s}$ increases the harvested energy and reduces EH waiting time, while improving the received SNR and thus the packet decoding probability. In addition, AAoI decreases rapidly at low $P_{\rm s}$ and gradually levels off as energy shortages become less frequent, whereas PLR continues to decrease with improved decoding reliability. Moreover, reducing $P_{\rm f}$ from 13 to 0 dBm increases FD errors, leading to higher AAoI and PLR. Conversely, reducing $R_{\rm b}$ from 5 to 2 kbps lowers the decoding threshold, thereby reducing both AAoI and PLR.

To investigate the impacts of EA, RT, and imperfect FD on AAoI and PLR, the following six schemes are considered. All schemes operate in the full-reflection mode during BT. For the No-EA schemes, energy harvested in one EH block can be used for transmission in the immediately following block but cannot be accumulated across consecutive EH blocks.
\begin{enumerate}
  \item \textbf{No-EA \& No-RT:} No-EA without RT.
  \item \textbf{No-EA \& RT \&  Perfect FD:} No-EA with RT and perfect FD.
  \item \textbf{No-EA \& RT \&  Imperfect FD:} No-EA with RT and imperfect FD.
  \item \textbf{EA \& No-RT:} Cross-block EA without RT.
  \item \textbf{EA \& RT \&  Perfect FD:} Cross-block EA with RT and perfect FD.
  \item \textbf{Proposed (EA \& RT \& Imperfect FD):} Cross-block EA with RT and imperfect FD.
\end{enumerate}
\begin{figure}
  \centering
  \includegraphics[width=0.63\linewidth]{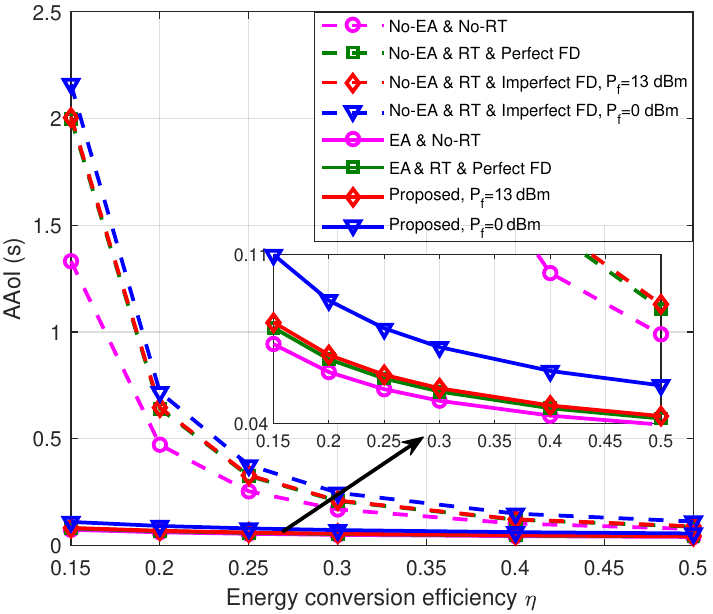}
    \vspace{-1em}
  \caption{Impact of EA, RT, and imperfect FD on AAoI.}\label{Fig5}
  \vspace{-1em}
\end{figure}
\begin{figure}
  \centering
  \includegraphics[width=0.67\linewidth]{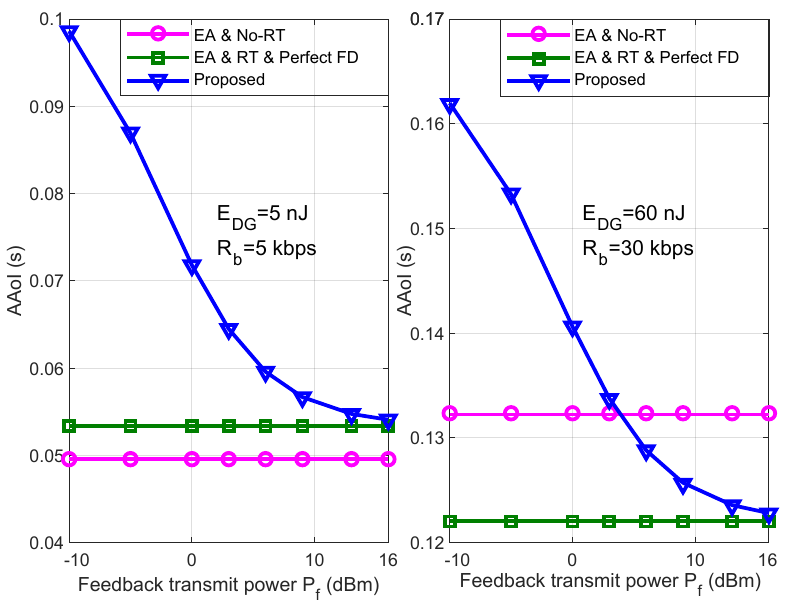}
    \vspace{-1em}
  \caption{Impact of feedback transmit power on AAoI under different settings.}\label{Fig6}
  \vspace{-1em}
\end{figure}
For a fair comparison, all schemes with RT incur the same EC for FD, regardless of whether the feedback is perfect or imperfect. In contrast, the No-RT schemes incur no EC for FD. All other common system parameters are kept identical.

Fig. \ref{Fig5} compares the AAoI of these schemes versus the energy conversion efficiency $\eta$. As $\eta$ increases, the AAoI of all schemes decreases due to increased harvested energy and reduced EH waiting time. Moreover, cross-block EA substantially reduces AAoI compared with no-EA, particularly at low $\eta$. This advantage gradually diminishes as $\eta$ increases, since a single block becomes more likely to provide sufficient energy for transmission, highlighting the importance of EA under energy-limited conditions.
Under the considered parameters, RT results in a higher AAoI than no-RT, even with perfect FD. This is because the age penalty of retransmitting old packets outweighs the reduction in EH waiting time achieved by avoiding repeated DG. Nevertheless, the AAoI gap between RT and no-RT is considerably smaller with EA, suggesting that EA alleviates the additional energy-related waiting associated with RT. Moreover, imperfect FD increases AAoI due to erroneous RT decisions, with a more pronounced degradation at lower feedback transmit power $P_{\rm f}$.

Although PLR is not plotted in Fig. \ref{Fig5}, the PLR under the proposed protocol follows \eqref{34A}. Without RT, PLR equals $P_{\rm out}$, whereas RT with perfect FD achieves zero PLR since failed packets are retransmitted until successful delivery. Under imperfect FD, PLR lies between zero and $P_{\rm out}$ and increases as $P_{\rm f}$ decreases due to more frequent NACK-to-ACK detection errors. Moreover, EA does not affect PLR under the same transmission and feedback conditions, since insufficient energy only delays transmission attempts rather than causing packet loss.
Overall, RT improves transmission reliability but increases AAoI under the considered conditions. The following results investigate whether this trade-off persists under different conditions and how imperfect FD influences it.

Fig. \ref{Fig6} compares the AAoI of the ``EA \& No-RT'', ``EA \& RT \& Perfect FD'', and proposed schemes under two parameter settings. As the feedback transmit power $P_{\rm f}$ increases, the AAoI of the proposed scheme decreases and approaches that under perfect FD. However, the relative AAoI performance of RT and no-RT differs between the two cases.
For $E_{\rm {DG}}=5$ nJ and $R_{\rm b}=5$ kbps, no-RT achieves a lower AAoI than both RT schemes.
Although RT avoids the additional DG energy incurred by no-RT after transmission failures, the low $E_{\rm DG}$ and high decoding success probability at low $R_{\rm b}$ limit the resulting reduction in EH waiting time, making it insufficient to offset the age penalty of retransmitting old packets.
In contrast, for $E_{\rm DG}=60$ nJ and $R_{\rm b}=30$ kbps, RT with perfect FD  achieves a lower AAoI than no-RT, while the proposed scheme shifts from a higher to a lower AAoI than no-RT as $P_{\rm f}$ increases.
In this case, the higher decoding failure probability and DG energy make repeated DG in no-RT more energy-intensive, potentially prolonging EH waiting time. RT avoids these additional costs, while improved feedback reliability reduces erroneous RT, allowing the savings in EH waiting time to outweigh the age penalty of retransmitting old packets.

\begin{figure}
  \centering
  \includegraphics[width=0.65\linewidth]{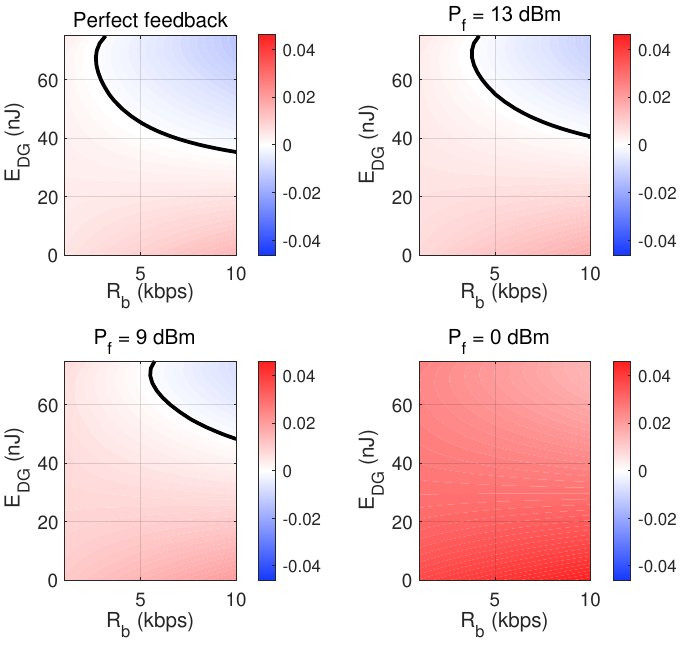}
    \vspace{-1em}
  \caption{AAoI performance regions of RT and No-RT schemes under different feedback conditions.}\label{Fig7}
  \vspace{-1em}
\end{figure}
\begin{figure}
  \centering
  \includegraphics[width=0.65\linewidth]{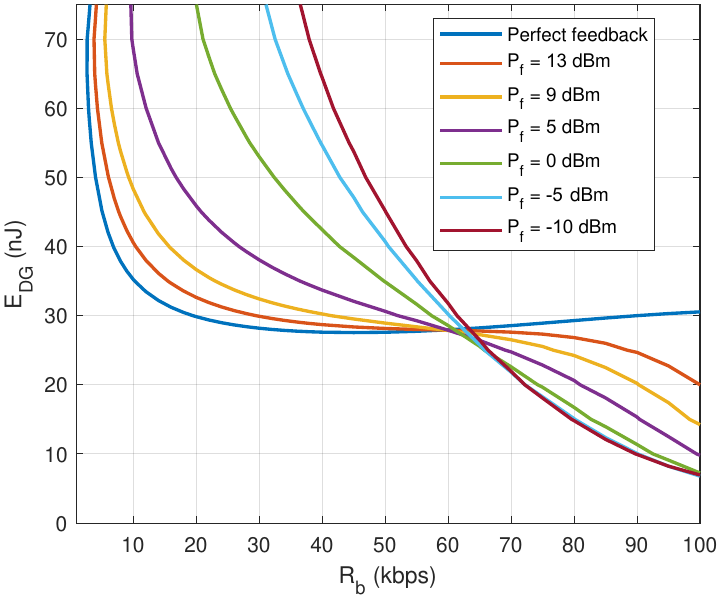}
    \vspace{-1em}
  \caption{Critical DG energy for equal AAoI between RT and No-RT schemes versus DT rate under different feedback conditions.}\label{Fig8}
  \vspace{-1em}
\end{figure}

To further identify the conditions under which RT achieves an AAoI advantage, Fig. \ref{Fig7} plots the AAoI difference $\Delta_{\rm AAoI}=\mathrm{AAoI}_{\rm RT}-\mathrm{AAoI}_{\rm No-RT}$ against $E_{\rm DG}$ and $R_{\rm b}$ under different feedback conditions. The color bars represent $\Delta_{\rm AAoI}$, with blue and red regions indicating lower AAoI for RT and no-RT, respectively, while the black contours correspond to $\Delta_{\rm AAoI}=0$. Under perfect FD, RT becomes more favorable as $E_{\rm DG}$ and $R_{\rm b}$ increase, since avoiding repeated DG can yield greater savings in EC and EH waiting time. However, as $P_{\rm f}$ decreases, the AAoI-favorable region of RT gradually shrinks, with the critical boundary shifting toward higher $E_{\rm DG}$ for a given $R_{\rm b}$. At $P_{\rm f}=0$ dBm, RT provides no AAoI advantage within the considered parameter range, indicating that feedback errors can offset the benefits of RT in reducing EH waiting time.

To examine whether this trend persists over a wider range of $R_{\rm b}$, Fig. \ref{Fig8} plots the critical $E_{\rm DG}^{}$ satisfying $\mathrm{AAoI}_{\rm RT}=\mathrm{AAoI}_{\rm No-RT}$. At relatively low $R_{\rm b}$, decreasing $P_{\rm f}$ increases $E_{\rm DG}^{}$, consistent with Fig. \ref{Fig7}. However, as $R_{\rm b}$ increases, $E_{\rm DG}^{}$ under perfect FD first decreases and then increases slightly, whereas it generally decreases under imperfect FD, with the critical boundaries intersecting at approximately $R_{\rm b}=65$ kbps. Beyond this point, their relative ordering reverses, indicating that poorer feedback reliability does not necessarily increase the critical $E_{\rm DG}^{}$ at higher $R_{\rm b}$. This behavior reflects the competing effects of DG and prolonged RT. As $R_{\rm b}$ increases, more frequent decoding failures increase the EC and potential EH waiting time associated with DG in no-RT, while also increasing the age penalty of repeatedly transmitting old packets in RT. Under perfect FD, prolonged RT may eventually offset the energy-saving benefits of RT, explaining the slight increase in $E_{\rm DG}^{}$ at high $R_{\rm b}$. Under imperfect FD, NACK-to-ACK detection errors may terminate RT of old packets prematurely, potentially reducing their age penalty at the expense of increased packet loss. This may partly explain the reversal of the critical boundaries. Overall, these results reveal that the AAoI advantage of RT depends jointly on DG energy, DT rate, and feedback reliability.

\begin{figure}
  \centering
  \includegraphics[width=0.7\linewidth]{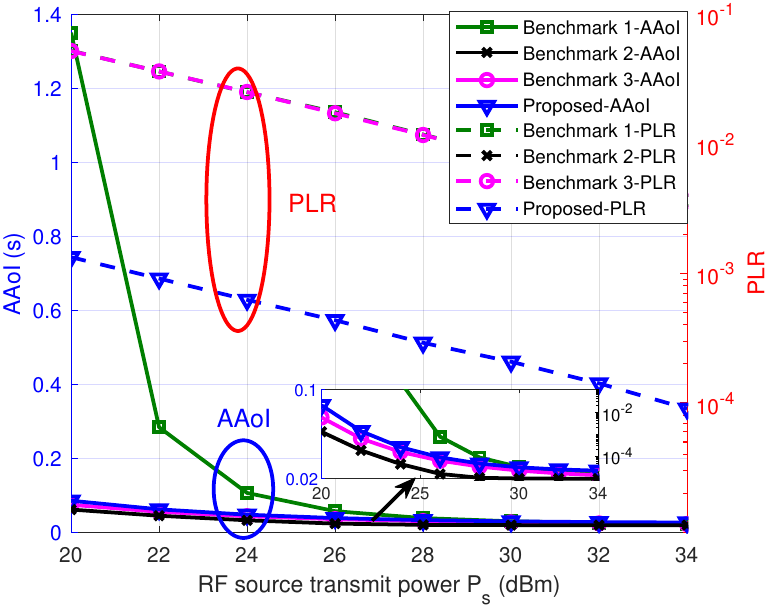}
    \vspace{-1em}
  \caption{AAoI and PLR comparison with existing benchmarks.}\label{Fig9}
  \vspace{-1em}
\end{figure}

Fig. \ref{Fig9} compares the AAoI and PLR of the proposed scheme with three benchmarks in existing works.
\begin{enumerate}
\item \textbf{Benchmark 1:} A conventional per-block energy management scheme without cross-block EA and RT \cite{10197252,11365549,11514149,10702441,11175015,10643608,9866050,9813592,10778600}, where the optimal PRC is determined subject to the energy causality constraint.

\item \textbf{Benchmark 2:} A cross-block energy management scheme without RT \cite{11482890}, where the optimal PRC is determined subject to both QoS and energy causality constraints, relying on accurate CSI of the BD$\to$R link to ensure successful decoding.

\item \textbf{Benchmark 3:} A cross-block energy management scheme without RT, where the optimal PRC is determined solely by the energy causality constraint, without requiring accurate CSI of the BD$\to$R link.
\end{enumerate}

Benchmark 1 exhibits a significantly higher AAoI, particularly at low $P_{\rm s}$, since the harvested energy cannot be accumulated. In contrast, the other schemes substantially reduce EH waiting time through EA, with the AAoI gap gradually narrowing as $P_{\rm s}$ increases. Among them, Benchmark 2 achieves the lowest AAoI, as the BD selects the minimum PRC satisfying the QoS requirement and transmits only when the energy causality constraint is also met. This avoids unsuccessful transmissions while allowing more incident RF power to be harvested, thereby reducing energy waste and EH waiting time between successful updates.
The proposed scheme exhibits a slightly higher AAoI than Benchmark 2 and 3 under the considered parameters, mainly due to the age penalty associated with retransmitting old packets and erroneous feedback decisions.

Regarding PLR, Benchmarks 1 and 3 exhibit nearly identical values, indicating that cross-block EA alone does not significantly affect packet delivery reliability under the considered conditions, despite its substantial AAoI improvement. In contrast, the proposed scheme achieves a much lower PLR than these two benchmarks through RT, with only a slight increase in AAoI relative to Benchmark 3. Benchmark 2 achieves zero PLR under its ideal CSI and QoS assumptions, since packets are generated and transmitted only when successful decoding is guaranteed. Although the proposed scheme exhibits a slightly higher AAoI and a low but nonzero PLR compared with Benchmark 2, it avoids the need for accurate instantaneous CSI feedback of the BD$\to$R link while accounting for practical imperfect FD, offering a more practical alternative under non-ideal operating conditions.

\section{Conclusions}\label{sec:Conclusions}
In this paper, we have proposed a threshold-triggered EA-then-act protocol for BackCom systems with RT under imperfect FD.
An embedded Markov chain was developed to characterize the steady-state distribution of stored energy, and analytical expressions for AAoI and PLR were derived and validated through simulations.
The results provide several important insights.
First, EA substantially decreases AAoI, particularly under energy-limited conditions, while having no impact on PLR.
Second, although RT improves transmission reliability, it may either improve or degrade information freshness, depending on whether the reduction in EH waiting time outweighs the age penalty of retransmitting old packets, which is influenced by DG energy, DT rate, and feedback power.
Third, imperfect FD alters the AAoI-favorable region of RT, as lower feedback power increases the critical DG energy required for RT to outperform no-RT at low DT rates, whereas the opposite trend emerges at higher DT rates.
Finally, the proposed protocol substantially decreases AAoI and PLR over conventional per-block energy management, while achieving an AAoI close to that of the QoS-aware cross-block EA scheme and a low PLR without requiring accurate CSI.

\section*{Appendix A}
\label{app:proof1}
The proof is divided into two cases according to whether the initial stored energy
$e_a$ satisfies the energy threshold $C$.

When $e_a\geq C$, the stored energy is sufficient to support the transmission
operation, and hence no EH is required. The BD directly performs one transmission operation with EC $C$. Therefore, the remaining continuous energy after transmission is given by $e_a-C$. After applying the downward quantization function $\xi(\cdot)$, the final discrete energy state is $\xi(e_a-C)$. Therefore, $P_{ab}^{(C)}=\mathbf{1}\{e_b=\xi(e_a-C)\}$, where $\mathbf{1}\{\cdot\}$ denotes the indicator function.

When $e_a<C$, the BD first performs EH until its stored energy becomes sufficient to support the transmission operation. Define the required
additional harvested energy as $c_a^{(C)}
=C-e_a>0$. Let $S_n=\sum_{i=1}^{n}E_{\rm H}(i)$ denote the cumulative harvested energy after $n$ EH blocks, where $E_{\rm H}(i)=\eta P_{\rm s}|f(i)|^2T_b$ is the harvested energy in the $i$-th EH block, following $E_{\rm H}(i)\sim\mathrm{exp}(\lambda_H)$, ${\lambda _H} \!=\! {{{\lambda _f}} \mathord{\left/
 {\vphantom {{{\lambda _f}} {\left( {\eta P{T_b}} \right)}}} \right.
 \kern-\nulldelimiterspace} {\left( {\eta P_{\rm s}{T_b}} \right)}}$ due to $|f(i)|^2\sim\mathrm{exp}(\lambda_f)$.
Then, the minimum
number of EH blocks required to reach the threshold is given by $N_C(e_a)=\inf\left\{n\geq1:S_n\geq c_a^{(C)}
\right\}$.

The excess harvested energy beyond the required energy deficit is defined as
$O
=
S_{N_C(e_a)}
-
c_a^{(C)}$.
Immediately before the last EH block, the remaining energy required to reach
the threshold is
$r
=
c_a^{(C)}
-
S_{N_C(e_a)-1}$.
Since the $N_C(e_a)$-th EH block is the first one that makes the accumulated
harvested energy reach or exceed $c_a^{(C)}$, it necessarily satisfies
$E_{\rm H}(N_C(e_a))
\geq r$.
Accordingly, the overshoot can be expressed as
$O
=
E_{\rm H}(N_C(e_a))-r$.
By the memoryless property of the exponential distribution, for any
$x\geq0$,
\begin{align}
\Pr
\left\{
E_{\rm H}(N_C(e_a))\!-\!r\!>\!x
\,\middle|\,
E_{\rm H}(N_C(e_a))\!\geq\! r
\right\}\!=\!
e^{-\lambda_Hx}.
\end{align}
Therefore, the conditional distribution of the overshoot does not depend on
the remaining energy $r$, and
$O
\sim
\mathrm{exp}(\lambda_H)$.

After the energy threshold $C$ is reached, the stored energy of the BD before the transmission operation can be expressed as $B_{\rm th}=\min\{B_{\max},C+O\}$, where the truncation operation is imposed due to the finite battery capacity. Then, after performing one transmission attempt with EC $C$, the remaining continuous stored energy is given by $R_C=\min\{B_{\max}-C,O\}$.
The final discrete energy state is obtained by applying the quantization
function $\xi(\cdot)$ to $R_C$. Therefore,
$P_{ab}^{(C)}
=
\Pr\{\xi(R_C)=e_b\}$.

To derive the closed-form expression of $P_{ab}^{(C)}$, define the maximum reachable discrete energy-state index after consuming energy
$C$ as
$Q_C
=
\min
\left\{
\left\lfloor
\frac{B_{\max}-C}{\Delta}
\right\rfloor,
Q
\right\}$.

For $0\leq b<Q_C$, the quantized energy state $s_b$ corresponds to the
continuous energy interval $[e_b,e_{b+1})$. Hence,
\begin{equation}\label{37A}
P_{ab}^{(C)}
=
\Pr\{e_b\leq O<e_{b+1}\}
=
e^{-\lambda_H e_b}
-
e^{-\lambda_H e_{b+1}}.
\end{equation}

For $b=Q_C$, the corresponding state represents the maximum reachable
discrete energy state after the transmission operation. Due to the finite
battery capacity, this state includes two cases: one is that the remaining
energy falls into the interval $[e_{Q_C},B_{\max}-C)$, and the other is that the
remaining energy exceeds $B_{\max}-C$ and is truncated by $B_{\max}$.
Therefore,
\begin{align}\notag\label{38A}
P_{ab}^{(C)}
&=
\Pr\{e_{Q_C}\leq O<B_{\max}-C\}
+
\Pr\{O\geq B_{\max}-C\}\\
&=
e^{-\lambda_H e_{Q_C}}.
\end{align}

For $b>Q_C$, the corresponding energy states cannot be reached after one
transmission attempt because the remaining stored energy is upper bounded by
$B_{\max}-C$. Thus, $P_{ab}^{(C)}=0$.


\section*{Appendix B}
\label{app:proof2}

When $e_a\geq C$, $N_C(e_a)=0$.
Therefore, the total number of blocks required
for completing one transmission attempt is $Z_C(e_a)=N_C(e_a)+1=1$.
Therefore, the first- and second-order joint moments are given by, respectively
\begin{equation}\notag
\begin{aligned}
T_{ab}^{(C)}
=
\mathbb{E}
\left[
Z_C(e_a)
\mathbf{1}_{\{e_a\rightarrow e_b\}}
\right]
=
\mathbb{E}
\left[
\mathbf{1}_{\{e_a\rightarrow e_b\}}
\right]
=
P_{ab}^{(C)},
\end{aligned}
\end{equation}
\begin{equation}\label{39A}
\begin{aligned}
U_{ab}^{(C)}
=
\mathbb{E}
\left[
Z_C^2(e_a)
\mathbf{1}_{\{e_a\rightarrow e_b\}}
\right]
=
P_{ab}^{(C)}.
\end{aligned}
\end{equation}

When $e_a<C$, the minimum number of EH blocks required to reach the threshold is
$N_C(e_a)
=
\inf
\left\{
n\geq1:
S_n\geq c_a^{(C)}
\right\}$,
where
$S_n=\sum_{i=1}^{n}E_{\rm H}(i)$ and $c_a^{(C)}=C-e_a$.
Since $E_{\rm H}(i)\sim\mathrm{exp}(\lambda_H)$, $S_n$ can be interpreted as the arrival epochs of a Poisson process with rate $\lambda_H$.
Let
$M\left(c_a^{(C)}\right)
=
\max
\left\{
n:
S_n<c_a^{(C)}
\right\}$
denote the number of arrivals before the energy deficit $c_a^{(C)}$ is
reached.
Then,
$M\left(c_a^{(C)}\right)
\sim
\mathrm{Poisson}\left(\mu_a^{(C)}\right)$,
where
$\mu_a^{(C)}
=
\lambda_H c_a^{(C)}
$.

Since the next EH block after these
$M\left(c_a^{(C)}\right)$ arrivals is the first one that makes the accumulated
harvested energy reach or exceed $c_a^{(C)}$, we have
$N_C(e_a)
=
M\left(c_a^{(C)}\right)+1$.
The overshoot beyond the
required energy deficit is
$O
=
S_{N_C(e_a)}
-
c_a^{(C)}$.
For the exponential EH process, the residual distance from the fixed level
$c_a^{(C)}$ to the next arrival is exponentially distributed with parameter
$\lambda_H$. Moreover, by the independent-increment property of the associated
Poisson process,
$M\left(c_a^{(C)}\right)
\perp
O$.
Therefore,
$N_C(e_a)
\perp
O$.

After consuming energy $C$ for the transmission operation, the remaining
continuous energy is
$R_C=
\min
\left\{
B_{\max}-C,
O
\right\}$.
Hence, the final discretized energy state $e_b=\xi(R_C)$ depends only on the
overshoot $O$. Since $N_C(e_a)$ and $O$ are independent for
the given initial energy level $e_a$, it follows that
$N_C(e_a)
\perp
\mathbf{1}_{\{e_a\rightarrow e_b\}}$.

Since $M\left(c_a^{(C)}\right)
\sim
\mathrm{Poisson}\left(\mu_a^{(C)}\right)$, the first- and second-order moments of
$M\left(c_a^{(C)}\right)$ are given by
\begin{equation}\label{40A}
\mathbb{E}\!
\left[
M\left(c_a^{(C)}\right)
\right]
\!=\!
\mu_a^{(C)},
\mathbb{E}\!
\left[
M^2\left(c_a^{(C)}\right)
\right]
\!=\!
\left(\mu_a^{(C)}\right)^2
\!+\!
\mu_a^{(C)}.
\end{equation}

Due to $Z_C(e_a)=N_C(e_a)+1=M\left(c_a^{(C)}\right)+2$, the first- and second-order joint moments can be computed by, respectively
\begin{align}\notag
T_{ab}^{(C)}
&=
\mathbb{E}
\left[
Z_C(e_a)
\mathbf{1}_{\{e_a\rightarrow e_b\}}
\right]
=
\mathbb{E}
\left[
Z_C(e_a)
\right]
P_{ab}^{(C)}
\\\label{41A}
&\!\!=\!\!
\left(
\mathbb{E}[M\left(c_a^{(C)}\right)]+2
\right)
P_{ab}^{(C)}
\!\!=\!\!
\left(
\mu_a^{(C)}+2
\right)
P_{ab}^{(C)}.
\end{align}
\vspace{-1em}
\begin{align}\notag
U_{ab}^{(C)}
&=
\mathbb{E}
\left[
Z_C^2(e_a)
\mathbf{1}_{\{e_a\rightarrow e_b\}}
\right]
=
\mathbb{E}
\left[
\left(
M\left(c_a^{(C)}\right)+2
\right)^2
\right]
P_{ab}^{(C)}
\\\label{42A}
&=
\left[
\left(\mu_a^{(C)}\right)^2
+
5\mu_a^{(C)}
+
4
\right]
P_{ab}^{(C)}.
\end{align}

\vspace{-1em}

\ifCLASSOPTIONcaptionsoff
  \newpage
\fi
\bibliographystyle{IEEEtran}
\bibliography{refa}

\end{document}